\documentclass{aa}

\usepackage{graphicx}
\usepackage{txfonts}
\usepackage[colorlinks=true,linkcolor=blue,citecolor=blue,filecolor=blue,urlcolor=blue]{hyperref}
\newcommand{\linkorcid}[1]{%
  \href{https://orcid.org/#1}{\protect\includegraphics[width=8pt]{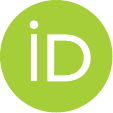}}%
}
\usepackage[babel]{csquotes}

\usepackage{amsmath, mathtools, amssymb}
\usepackage[locale = US]{siunitx}

\usepackage[percent]{overpic}

\usepackage[acronym]{glossaries}
\glsdisablehyper
\makeglossaries
\newacronym{SDSS}{SDSS}{Sloan Digital Sky Survey}
\newacronym{HSC}{HSC}{Hyper Suprime-Cam}
\newacronym[longplural={Multi Layer Perceptrons}, shortplural={MLPs}]{MLP}{MLP}{Multi Layer Perceptron}
\newacronym[longplural={Rectified Linear Units}, shortplural={ReLUs}]{ReLU}{ReLU}{Rectified Linear Unit}
\newacronym{CGM}{CGM}{Classification-Guided Module}
\newacronym{ADAMW}{ADAMW}{Adaptive Moment Estimation with decoupled Weight decay}
\newacronym{FPR}{FPR}{False-Positive Rate}
\newacronym{FNR}{FNR}{False-Negative Rate}
\newacronym{TPR}{TPR}{True-Positive Rate}
\newacronym{TNR}{TNR}{True-Negative Rate}
\newacronym{NED}{NED}{NASA/IPAC Extragalactic Database}
\newacronym{cCN}{cCN}{compact Convolutional Neural Network}
\begin{document}

   \title{Searching for low-surface-brightness galaxies with compact neural networks}

   \subtitle{A parameter-efficient approach to first-pass selection of diffuse galaxy in HSC-SSP imaging}

   \author{G\"{u}nther K. Heemann\inst{1}\linkorcid{0009-0001-8073-3428}
          \and Henri Cecatka\inst{1}\linkorcid{0009-0009-6372-451X}
          \and Dominik J. Bomans\inst{1}\linkorcid{0000-0001-5126-5365}
          }

   \institute{Ruhr University Bochum, Faculty of Physics and Astronomy, Astronomical Institute (AIRUB), Universitätsstraße 150, 44801 Bochum, Germany\\
              \email{heemann@astro.ruhr-uni-bochum.de}
             }


 
\abstract
{Low-surface-brightness galaxies (LSBGs) trace a diffuse and observationally 
challenging regime of galaxy formation that remains highly susceptible to selection 
effects in optical surveys.}
{We present a hybrid LSBG detection pipeline that integrates compact convolutional 
neural networks (cCNs) as a morphological validation stage within a multi-step 
detection framework applied to HSC-SSP imaging.}
{Permissive low-threshold source detection with \textsl{SEP} is followed by 
cCN-based morphological filtering, physically motivated consistency checks, and 
parametric surface-brightness modelling with \textsl{galfitm}. The cCN is 
deliberately constrained to ${\sim}10^4$ trainable parameters to match the spatial 
scales of diffuse emission and ensure computational efficiency at survey scale.}
{The cCN reduces the initial detection set of ${\sim}77\times10^6$ objects by more than four orders of magnitude while recovering essentially all literature LSBGs that pass the detection stage. The resulting Gold catalog of 5156 candidates occupies the characteristic loci of LSBGs in colour--colour and structural parameter space, in close agreement with previously published HSC-SSP based samples. Cross-matching with SDSS and DESI spectroscopic redshifts confirms that the pipeline identifies genuine nearby diffuse galaxies.}
{Compact neural networks provide a robust and computationally scalable morphological 
filter for LSBG searches, effectively bridging classical detection pipelines and 
large general-purpose deep architectures. The pipeline is detection-limited rather 
than selection-limited, identifying the initial source extraction stage as the 
primary target for future improvement.}

   \keywords{galaxies: dwarf --
              galaxies: photometry --
              galaxies: structure --
              methods: data analysis --
              methods: machine learning --
              techniques: image processing
             }

   \maketitle
%

\section{Introduction}

Low-surface-brightness galaxies (LSBGs) represent a significant yet underexplored 
component of the galaxy population. Early work already demonstrated that galaxies 
with central surface brightnesses comparable to or fainter than the night sky are 
intrinsically difficult to detect, leading to strong observational selection effects 
against diffuse systems \citep{Disney1976}. As a consequence, classical optical 
surveys systematically underrepresented low-luminosity and extended galaxies, biasing 
early empirical views of galaxy demographics and scaling relations 
\citep[e.g.,][]{Impey1997, Blanton2005}.

Subsequent studies established that LSBGs span a broad range of physical properties 
and environments, from isolated field galaxies to dense cluster regions 
\citep[e.g.,][]{Bothun1987, Impey1996}. In the context of the $\Lambda$ cold dark 
matter ($\Lambda$CDM) paradigm, such diffuse systems are a natural outcome of galaxy 
formation in dark matter halos with high angular momentum or inefficient star 
formation \citep{DalCanton1997}. Indeed, several long-standing challenges 
to $\Lambda$CDM on small scales were first identified in the low-luminosity and 
low-surface-brightness regime \citep[e.g.,][]{Bullock2017, Kauffmann1993, Boylan-Kolchin2011, Moore1999}.

Cosmological hydrodynamical simulations have significantly clarified the physical 
nature of low-surface-brightness galaxies within the $\Lambda$CDM framework. 
Modern large-volume simulations such as \textsl{Horizon-AGN}, textsl{IllustrisTNG}, and \textsl{EAGLE}  indicate that LSBGs form naturally within the $\Lambda$CDM framework, although the dominant channel remains debated. Some studies find that diffuse systems preferentially occupy high-spin dark matter halos \citep{Perez-Montano2022, Stoppacher2025}, whereas others recover comparable specific angular momenta for low- and high-surface-brightness systems and instead attribute the diffuse morphology to assembly-history effects — feedback-driven expansion, tidal perturbations, and ram-pressure stripping — that redistribute stellar mass to large radii \citep{Martin2019, Jackson2021}. Across these models, LSBGs nonetheless emerge as genuinely extended systems with reduced central stellar densities, representing a distinct population rather than observational artifacts. This divergence can be traced to early cosmic times: in the EAGLE simulation the specific angular momenta of the two populations begin to diverge at z\textasciitilde 5-7, while star-formation activity and large-scale environment play only a minor role in shaping the low-surface-brightness features \citep{Stoppacher2025}.
However, translating this theoretical picture into observational constraints requires 
detection strategies that reliably separate extended low-contrast galaxies from 
background fluctuations, scattered light, and instrumental residuals 
\citep{Mihos2015, Greco2018}. This is not merely a technical challenge but a 
prerequisite for confronting theoretical predictions with observational data. 
Large-area surveys such as the Sloan Digital Sky Survey (SDSS) enabled statistical samples of LSBGs with central surface brightnesses down to $\mu0(B) \sim 23.5$~mag~arcsec$^{-2}$ and studies of their large-scale environments, although fainter populations remained largely inaccessible \citep[e.g.,][]{Zhong2008, Rosenbaum2009}.

The advent of modern wide-field optical imaging surveys has marked a turning point 
for LSBG studies. In particular, the Hyper Suprime-Cam Subaru Strategic Program 
(HSC-SSP) combines exceptional depth, image quality, and survey area, enabling 
statistical studies of diffuse galaxies over a wide range of environments 
\citep{Aihara2018, Aihara2022}. Using HSC-SSP data, \citet{Greco2018} demonstrated that 
galaxies with mean effective surface brightnesses of $\bar{\mu}_{\mathrm{eff}} \gtrsim 
24.3$~mag~arcsec$^{-2}$ can be systematically detected when pipelines are explicitly 
optimized for low surface brightness emission. 
The availability of other deep, wide-field datasets including the Dark Energy Survey (DES) and the Kilo Degree Survey (KiDS) \citep{DES, KIDS} has enabled further studies, many of which have utilised machine learning to identify LSBG candidates \citep[e.g.,][]{Tanoglidis2021, Su2024LSBGnet:Galaxies, Thuruthipilly2025DESLearning, Su:2026:A&A}.

Throughout this work, we adopt an operational definition of low-surface-brightness 
galaxies consistent with \citet{Greco2018}. Specifically, we classify objects as 
LSBGs if they satisfy
\begin{equation}
\bar{\mu}_{\mathrm{eff},g} > 24.3~\mathrm{mag~arcsec^{-2}}
\end{equation}
and
\begin{equation}
r_{\mathrm{eff}} > 2.5'' ,
\end{equation}
where $\bar{\mu}_{\mathrm{eff},g}$ denotes the mean effective surface brightness in 
the $g$ band and $r_{\mathrm{eff}}$ is the circularized half-light radius derived 
from single-component Sérsic modeling. These criteria are purely observational and 
are not intended to imply a distinct physical galaxy class, but rather to define a 
regime in surface-brightness–size space that is traditionally associated with diffuse 
systems in wide-field optical surveys.

Nevertheless, the detection of LSBGs remains fundamentally limited by contamination. 
Standard source extraction algorithms such as \textsl{SExtractor} are not designed 
for low-surface-brightness emission and perform poorly when applied to diffuse 
systems. However, when operated at permissive detection thresholds, they can recover 
extended low-contrast sources, at the cost of introducing a large number of spurious 
detections, as background fluctuations, scattered light, galactic cirrus, and 
instrumental residuals produce morphologically similar signatures 
\citep[e.g.,][]{Mihos2015, Greco2018, Bertin1996}. Alternative approaches tailored 
to low-surface-brightness detection, such as \textsl{NoiseChisel}, mitigate some of 
these issues but still require careful validation of detected sources 
\citep{Akhlaghi2015}.

As a result, modern LSBG searches increasingly adopt multi-stage pipelines in which 
permissive detection is followed by morphological and physical validation 
\citep{Greco2018, Prole2019, Tanoglidis2021}. Within this framework, machine learning 
techniques provide a natural extension by enabling non-linear classification based on 
spatially resolved image information. Convolutional neural networks (CNNs) have been 
successfully applied to a variety of astronomical imaging tasks, including galaxy 
morphology classification and artifact rejection \citep[e.g.,][]{Dieleman2015, HuertasCompany2015, DominguezSanchez2018}.

However, many CNN architectures employed in astronomy are adapted from models developed for natural image classification and contain millions of trainable parameters \citep[][e.g.,]{Su2024LSBGnet:Galaxies, Liang2024AutomaticLearning}. For tasks involving small image cutouts and localized 
morphological decisions, such complexity is not always necessary and can introduce 
increased computational cost and susceptibility to overfitting \citep{Howard2017, Ntampaka2019}.
In this work, we investigate the use of compact convolutional neural networks (cCNs), a network architecture with $~10^3-10^5$ parameters as described in Section~\ref{subsec::cCN}, as a morphological filter tool for LSBG searches in HSC-SSP imaging data. By deliberately constraining model complexity to match the spatial scales and signal characteristics of diffuse galaxies, cCNs offer an efficient alternative to deeper architectures. The network is integrated as a selective filtering stage within a hybrid detection pipeline, complementing classical detection algorithms rather than replacing them.


\section{Convolutional neural network training}
\label{sec:cnn_training}

In this section we describe the properties of compact convolutional neural networks 
(\ref{subsec::ccns}), the construction of the positive and negative training samples 
(\ref{subsec:training_dataset}), the preprocessing of the HSC-SSP images and the 
generation of fixed-size network inputs (\ref{subsec:preprocessing_stacking}), the 
network architecture (\ref{subsec:cCN_Structure}), the training and hyperparameter 
optimisation strategy (\ref{subsec:training_protocol}), and the resulting performance 
on independent validation and test data (\ref{subsec:training_validation}).

\subsection{Compact Convolutional Neural Networks (cCNs)}
\label{subsec::ccns}

Image-based inference in astronomy now relies routinely on convolutional neural networks (CNNs), which have proved effective for classifying galaxy morphologies, characterising individual sources, and building catalogues at survey scale \citep[e.g.][]{Dieleman2015, HuertasCompany2015, DominguezSanchez2018}. In most cases, however, these networks inherit their design from architectures originally developed for generic natural-image recognition, with parameter counts reaching into the millions. Such capacity is rarely warranted for problems as constrained as classifying small cutouts from localised morphological cues, and it carries a cost in computational load and an increased risk of overfitting \citep{Chollet2017, Ntampaka2019}.

In this work, we refer to deliberately constrained lightweight architectures as compact convolutional neural networks (cCNs). These networks retain the essential inductive biases of convolution while explicitly limiting depth, channel width, and receptive-field growth. A key architectural element enabling this is depthwise-separable convolution (DSC), which factorizes standard convolutions into depthwise and pointwise operations, significantly reducing the number of trainable parameters while preserving expressive power \citep{Chollet2017, Howard2017}.

For low-surface-brightness science, this design philosophy is particularly well motivated. Discriminative signals in LSB galaxies are weak, extended, and easily confused with background fluctuations or reduction artifacts \citep{Greco2018, Tanoglidis2021}. By limiting sensitivity to misleading high-frequency features, cCNs naturally emphasize the coherent low-contrast structures characteristic of genuine diffuse galaxies. Within the hybrid pipeline described in 
Section~\ref{sec::Pipeline}, the cCN operates not as a blind source detector but as 
a selective filtering stage on the candidate list produced by \textsl{SEP}, 
prioritizing sample purity while keeping computational cost tractable at survey scale.

\subsection{Training dataset: positive and negative samples}
\label{subsec:training_dataset}

All training data are derived from HSC-SSP PDR3 imaging \citep{Aihara2022} and are 
square postage stamps centred on a given sky coordinate. We construct a binary 
classification dataset consisting of positive examples (diffuse galaxies consistent 
with the targeted LSBG regime) and negative examples (non-LSBG sources and 
empty-sky regions).

Positive examples are taken from the HSC-SSP LSBG catalogue of \citet{Greco2018}. For 
each catalogue entry, we extract a postage stamp centred on the catalogue 
coordinates. This provides a training set of diffuse, extended sources with 
surface-brightness properties that closely match our intended application domain.

The negative dataset is designed to reflect the dominant failure cases of permissive 
low-threshold source extraction, namely compact or structured sources as well as 
background fluctuations in apparently empty regions. For this purpose we use the 
public SuperBoRG catalogue \citep{Morishita2021SuperBoRG:Data} to obtain sky 
positions of sources detected in HST imaging that overlap with the HSC-SSP footprint, 
and extract the corresponding HSC-SSP postage stamps at these positions. In addition, we 
explicitly include negative examples from sky regions in the HSC-SSP images that are 
free of catalogue sources. These empty-sky stamps are generated by constructing a 
mask of all catalogue objects and the HSC-SSP detection mask, and extracting cutouts 
from the remaining unmasked regions. Masked pixels are permitted at the edges of a 
stamp, provided that the central region remains unmasked, ensuring that the cCN 
input is not dominated by missing data.

The combined dataset (LSBGs, random catalogue objects, and random empty-sky 
positions) comprises $\sim 10^4$ postage stamps. This combination is chosen to 
expose the network to a wide range of non-LSBG morphologies and backgrounds while 
maintaining a positive sample that is representative of the diffuse galaxy population.

\subsection{Image preprocessing, stacking, and network inputs}
\label{subsec:preprocessing_stacking}

Training and inference are performed on a single-channel detection image constructed 
from the HSC-SSP \textit{g}, \textit{r}, and \textit{i} bands. For each pixel position 
$(x,y)$ we form an inverse-variance weighted stack,
\begin{equation}
I_{\mathrm{stack}}(x,y)
=
\sum_{b=1}^{N_{\mathrm{bands}}}
\frac{I_b(x,y)}{\sigma_b^2(x,y)} \, ,
\label{eq:Stack}
\end{equation}
where $I_b(x,y)$ is the pixel value in band $b$ and $\sigma_b^2(x,y)$ denotes the 
corresponding variance. In this construction, pixels with large variance contribute 
less strongly to the detection image, which is advantageous for the low 
signal-to-noise regime relevant for diffuse emission.

All postage stamps used for training are generated from this stacked detection image 
and mapped to a fixed input size of $32\times 32$ pixels, matching the cCN input 
layer. This choice provides a compact representation that preserves the central 
morphology while controlling computational cost. The same preprocessing pipeline is 
applied during survey-scale inference to ensure that the network operates on inputs 
drawn from the same distribution as the training data.

\subsection{cCN Structure}
\label{subsec:cCN_Structure}

The network operates on single-channel $32\times 32$ pixel inputs constructed from 
the inverse-variance weighted $gri$ stack 
(Section~\ref{subsec:preprocessing_stacking}). It consists of three convolutional 
stages with interleaved max-pooling, followed by two fully connected layers and a 
single-logit output for binary classification. The detailed architecture is 
summarised in Table~\ref{tab:cCN_architecture}.

\begin{table*}[t]
\caption{Architecture of the compact CNN used in this work. Output shapes are given 
for an input of size $1\times 32\times 32$.}
\label{tab:cCN_architecture}
\centering
\begin{tabular}{lcccc}
\hline\hline
Layer & Kernel / Units & Output shape & Activation & Notes \\
\hline
Input & -- & $1\times 32\times 32$ & -- & stacked image \\
Conv2D & $3\times3$, 8 & $8\times 32\times 32$ & ReLU & stride 1, pad 1 \\
MaxPool & $2\times2$ & $8\times 16\times 16$ & -- & stride 2 \\
DSC & $3\times3$, 16 & $16\times 16\times 16$ & ReLU & depthwise + pointwise \\
MaxPool & $2\times2$ & $16\times 8\times 8$ & -- & stride 2 \\
DSC & $3\times3$, 32 & $32\times 8\times 8$ & ReLU & depthwise + pointwise \\
MaxPool & $2\times2$ & $32\times 4\times 4$ & -- & stride 2 \\
Flatten & -- & 512 & -- & $32\times4\times4$ \\
Fully connected & 16 & 16 & CELU & -- \\
Fully connected & 8 & 8 & CELU & -- \\
Output & 1 & 1 & logit & sigmoid at inference \\
\hline
\end{tabular}
\end{table*}

Depthwise-separable convolution (DSC) layers factorise a standard convolution into 
a channel-wise spatial convolution followed by a $1\times1$ pointwise convolution. 
Batch normalisation and ReLU activation are applied after each convolutional block. 
The total number of trainable parameters is $\sim10^4$, placing the model firmly in 
the compact regime relative to commonly used deep CNN architectures.

\subsubsection{Computational Efficiency of DSCs}
\label{subsubsec:cCN_Efficiency}

The use of DSCs provides substantial computational advantages over standard 
convolutions. For a single $32\times 32$ pixel image, the cCN requires approximately 
$175{,}240$ multiply-accumulate operations (MACs), compared to $671{,}880$ MACs for 
an equivalent network using standard convolutions — a reduction of $73.9\%$. The 
efficiency gains are primarily driven by the two DSC layers:
\begin{itemize}
    \item \textbf{DSC Stage 2} (8$\to$16 channels, $16\times16$ feature maps): 
    51,200 MACs vs. 294,912 MACs
    \item \textbf{DSC Stage 3} (16$\to$32 channels, $8\times8$ feature maps): 
    41,984 MACs vs. 294,912 MACs
\end{itemize}
This factorization achieves its efficiency through
\begin{equation}
\text{MACs}_{\text{DSC}} = H \times W \times K^2 \times C_{\text{in}} + 
H \times W \times C_{\text{in}} \times C_{\text{out}}
\end{equation}
compared to the standard convolution cost
\begin{equation}
\text{MACs}_{\text{Conv}} = H \times W \times K^2 \times C_{\text{in}} \times 
C_{\text{out}},
\end{equation}
where $H$ and $W$ are spatial dimensions, $K$ is kernel size, and 
$C_{\text{in}}/C_{\text{out}}$ are input/output channel counts. At survey scale, 
this translates directly to reduced memory requirements, lower power consumption, 
and faster inference times, all critical for morphological filtering of the 
${\sim}77\times10^6$ initial \textsl{SEP} detections processed by the pipeline.

\subsection{Training protocol and hyperparameter optimisation}
\label{subsec:training_protocol}

The dataset is randomly split into training ($70\,\%$), validation ($15\,\%$), and test ($15\,\%$) subsets, stratified with respect to both class label and postage stamp origin (LSBG, random catalogue object, random empty-sky region). This preserves the original distribution across subsets and reduces the risk that validation or test performance is biased by a different negative example composition.

To reduce overfitting, we apply standard regularisation and augmentation during training. Augmentation is performed through random spatial transformations including random cropping and flipping, improving robustness to small centring offsets and orientation changes. Weight decay is applied following the decoupled regularisation scheme of \citet{Loshchilov2019DecoupledRegularization}. Hyperparameters (learning rate, batch size, weight decay) are optimised using the \textsl{Optuna} framework \citep{Akiba2019Optuna:Framework} based on validation-set performance, with model selection emphasizing purity to reflect the downstream cost of false positives.

\subsection{Performance on independent test data}
\label{subsec:training_validation}

\begin{figure}
    \centering
    \includegraphics[width=\linewidth]{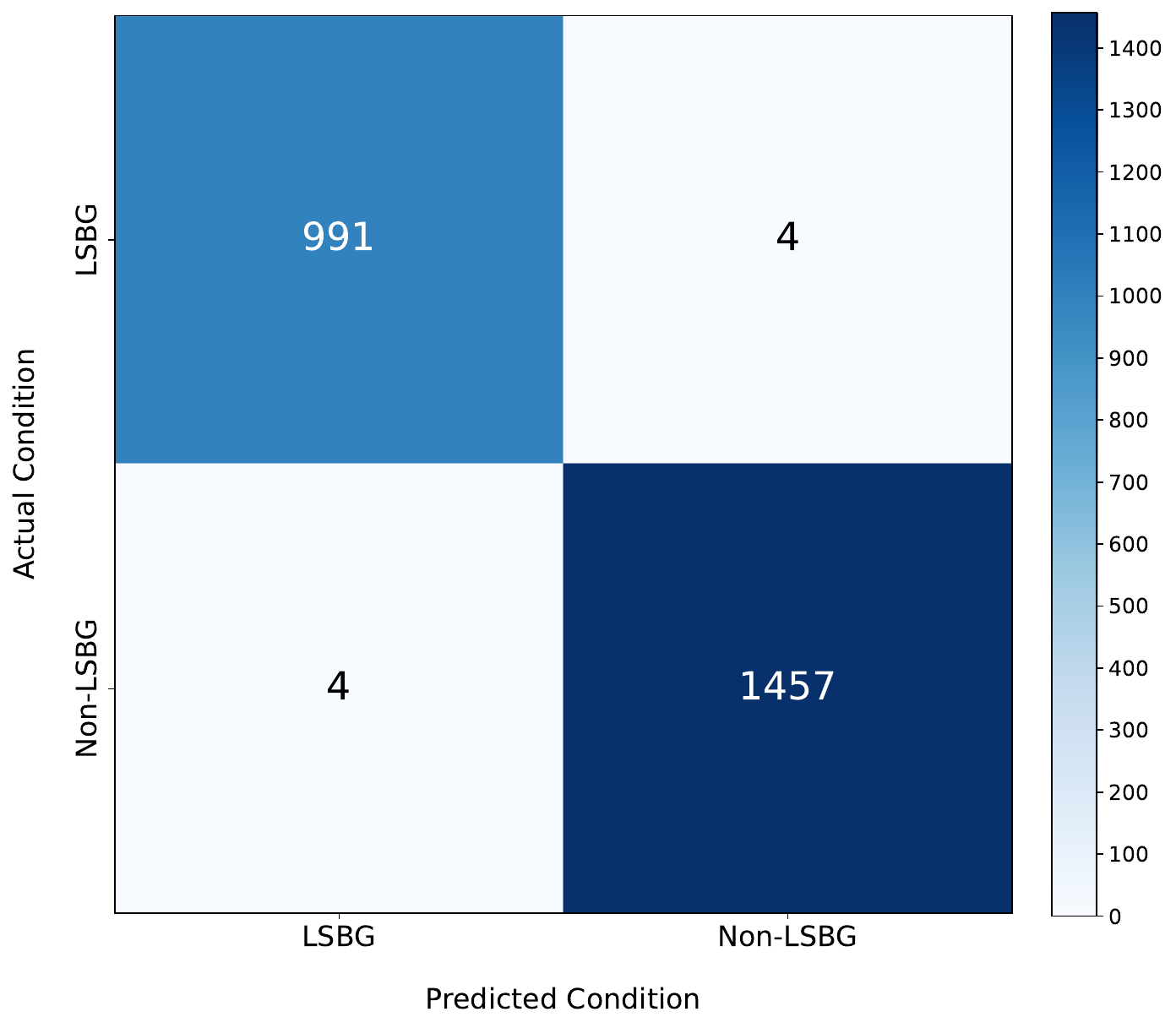}
    \caption{Confusion matrix for the independent test dataset.}
    \label{fig:confusion_matrix}
\end{figure}

The final model is evaluated on the held-out test subset, which is not used during training or hyperparameter optimisation. In addition to the data basis outlined above, postage stamps from \citet{Tanoglidis2021} are added to augment the positive sample. The resulting confusion matrix is shown in Fig.~\ref{fig:confusion_matrix}. The model achieves an F1 score of $99.60\,\%$ and a Matthews correlation coefficient of $99.30\,\%$, indicating near-perfect discrimination between diffuse galaxies and non-LSBG sources under controlled conditions. The survey-scale performance of the classifier within a realistic detection environment is evaluated separately in Section~\ref{sec:Crossmatch}.

\section{Pipeline}
\label{sec::Pipeline}

The detection pipeline proceeds in five sequential stages: permissive source 
detection with \textsl{SEP} (\ref{subsec::SEP}), morphological filtering with the cCN (\ref{subsec::cCN}), physically 
motivated pre-filters (\ref{subsec::Phys_Filter}), parametric surface-brightness modelling with 
\textsl{galfitm} (\ref{subsec::GALFITM}), and automated quality cuts to define the final candidate sample (\ref{sec:gold_catalog}). 
The object counts at each stage are summarised in Table~\ref{tab:pipeline_stages}.

\subsection{Source detection with SEP}
\label{subsec::SEP}

Initial source detection is performed on the inverse-variance weighted detection 
image $I_\mathrm{stack}$ (Eq.~\ref{eq:Stack}) constructed from the 
background-subtracted $g$-, $r$-, and $i$-band calexp images. A combined pixel mask 
is formed from the HSC-SSP mask planes of all three bands (excluding the detection 
mask itself) and passed to \textsl{SEP} \citep{Barbary2016}, which implements the 
core source extraction algorithm of \textsl{SExtractor} \citep{Bertin1996}.

A per-image background model is estimated using a mesh size of $128 \times 128$ 
pixels and a $3 \times 3$ median filter. The detection threshold is set to 
$\mathrm{THRESH\_SIGMA} = 0.6$ times the background RMS — an intentionally 
permissive value chosen to retain extremely faint and diffuse emission at the cost 
of a large number of false detections at this stage. Detected pixels are grouped 
into sources only if they form connected regions of at least $\mathrm{MINAREA} = 80$ 
contiguous pixels above the threshold. Source deblending is controlled by 
$\mathrm{DEBLEND\_CONT} = 0.001$ with 32 $\mathrm{DEBLEND_NTHRESH} = 32$, suppressing the 
fragmentation of diffuse galaxies while separating genuinely overlapping objects.

This stage yields ${\sim}76.98 \times 10^6$ detections across the processed 
footprint, reflecting the deliberately permissive detection strategy that 
prioritises completeness over purity.

\subsection{cCN filter}
\label{subsec::cCN}

The full \textsl{SEP} catalogue is evaluated by the cCN described in 
Section~\ref{sec:cnn_training}, reducing the candidate set by more than four orders 
of magnitude to ${\sim}59.7\times10^4$ candidates before computationally expensive steps are applied. For each detection, 
a $32 \times 32$ pixel cutout is extracted from $I_\mathrm{stack}$ and processed 
following the identical preprocessing described in 
Section~\ref{subsec:preprocessing_stacking}, ensuring consistency between training 
and survey-scale inference. The network produces a sigmoid-activated probability; 
detections with $p \geq 0.95$ are retained and forwarded to subsequent stages.

\subsection{Physical pre-filters}
\label{subsec::Phys_Filter}

All cCN-selected candidates are subjected to two physically motivated filters 
applied sequentially on the $g$-band calexp image, designed to remove contaminants while remaining agnostic to detailed morphological structure.

\subsubsection{Central surface-brightness filter}

For each candidate, the flux within a circular aperture of radius $r_\mathrm{ap} = 
2$ pixels is measured at the detection centroid using \texttt{sep.sum\_circle}. The 
aperture flux is converted to a central surface-brightness proxy $\mu_\mathrm{central}$ via
\begin{equation}
    \mu_\mathrm{central} = \mathrm{ZP} - 2.5\log_{10}\!\left(\frac{F_\mathrm{ap}}
    {\pi\,r_\mathrm{ap}^2\,p^2}\right),
\end{equation}
where $\mathrm{ZP} = 27.0$ is the photometric zeropoint and $p = 
0.167''\,\mathrm{pix}^{-1}$ is the HSC-SSP pixel scale. Objects outside the range 
$16 < \mu_\mathrm{central} < 30\ \mathrm{mag\,arcsec}^{-2}$ are rejected, removing 
saturated sources and strongly PSF-dominated objects at the bright end and 
background-dominated detections at the faint end.

\subsubsection{Ring-based extended-emission test}

As a morphological constraint independent of absolute surface-brightness 
normalisation, a ring-based test is applied to each candidate. Pixel values are 
evaluated on a narrow annulus centred on the detection centroid at angular radius 
$r_\mathrm{ring} = 2.5''$ (approximately 15 pixels), with a half-width of $0.75$ 
pixels. The reference intensity $I_\mathrm{core}$ is defined as the median pixel 
value within a circular aperture of radius $1$ pixel around the detection centre. 
A candidate is retained only if at least a fraction $f_\mathrm{min} = 0.05$ of the 
annulus pixels satisfies
\begin{equation}
    I_i \;\ge\; f_\mathrm{ring}\,I_\mathrm{core}, \qquad f_\mathrm{ring} = 0.15.
\end{equation}
Compact sources with steep radial profiles typically fail this test, whereas objects 
with extended low-surface-brightness emission at larger radii are preferentially 
retained.

The two physical pre-filters together reduce the candidate set from $592{,}739$ to 
$557{,}106$ ($6.0\%$ combined rejection), with $7{,}893$ removed by the 
$\mu_\mathrm{central}$ cut and a further $27{,}740$ by the ring test.

\subsection{Structural modeling with GALFITM}
\label{subsec::GALFITM}

All candidates passing the physical pre-filters are subjected to multi-band 
parametric surface-brightness modelling using \textsl{galfitm} 
\citep{Peng2002,Peng2010,pygalfitm_software,galfitm_software} via the \texttt{pygalfitm} interface. 
Modelling is performed simultaneously in the $g$, $r$, and $i$ bands to leverage 
colour information and improve parameter constraints for faint extended sources.

For each candidate, square image stamps of size $80 \times 80$ pixels 
(${\approx}13.4'' \times 13.4''$) are extracted from the calexp images of each 
band, centred on the detection coordinates. Candidates whose extraction region 
extends beyond the image boundaries are discarded. Uncertainty maps are derived from the variance planes of each calexp, where non-finite or non-positive variance 
pixels are replaced by the median uncertainty of valid pixels in the stamp. Pixel masks 
are formed from the HSC-SSP mask planes of each band in the same manner as before.

The PSF is extracted independently for each band from the HSC-SSP at the patch 
centre using \textsl{unagi} \citep{Huang2026unagi} and normalised to unit flux. If 
extraction fails for one or more bands, the best available PSF (priority: $i > r > 
g$) is substituted.

Each candidate is modelled with a single-component S\'ersic profile. Structural 
parameters (effective radius, S\'ersic index, axis ratio, position angle, and 
centroid) are constrained to be identical across bands, while magnitudes vary 
independently to capture colour information. Initial parameter values are derived 
from the \textsl{SEP} measurements, with the S\'ersic index initialised to $n = 1$. 
Fits are performed with a fixed zeropoint of $\mathrm{ZP} = 27.0$ mag and no sky 
refitting, subject to a timeout of $600\,\mathrm{s}$. Candidates that time out, 
fail to converge, or produce non-finite structural parameters are discarded.

From the best-fitting model, the effective surface brightness $\mu_\mathrm{eff}$ 
and central surface brightness $\mu_0$ are computed in the $g$ band as
\begin{align}
    I_e &= \frac{10^{-0.4(m_g - \mathrm{ZP})}}{\pi\,r_e^2\,p^2\,q}, \notag\\
    \mu_\mathrm{eff} &= \mathrm{ZP} - 2.5\log_{10}(I_e), \notag\\
    \mu_0 &= \mathrm{ZP} - 2.5\log_{10}(I_e\,e^{b_n}) \notag
\end{align}
where $r_e$ is the effective radius in pixels, $q$ the fitted axis ratio, and $b_n$ 
the standard S\'ersic shape parameter. Sky coordinates are derived exclusively from 
the fitted centroid converted via the $i$-band WCS.

\textsl{galfitm} modelling reduces the candidate count from $557{,}106$ to 
$541{,}256$ ($2.8\%$ rejection), primarily due to convergence failures and stamp 
boundary violations.

\subsection{Construction of the Gold catalog}
\label{sec:gold_catalog}

From the converged \textsl{galfitm} fits, a high-purity subsample is defined by 
automated quality cuts and colour-based classification, following a deliberately 
conservative strategy aimed at maximising reliability rather than completeness 
\citep[e.g.][]{Greco2018, Tanoglidis2021}.

Candidates are required to satisfy
\begin{center}
\(\displaystyle
\begin{aligned}
    24.3 \le &\mu_{\mathrm{eff},g} < 30.0\ \mathrm{mag\,arcsec}^{-2}, \\
    2.5'' \le &R_{e}               \le 90.0'', \\
    0.5 < &n                       \le 5.0, \\
    &q                             \ge 0.2, \\
    &\frac{\chi^2}{\nu}            \le 3.0
\end{aligned}
\)
\end{center}
Objects with $\mu_{\mathrm{eff},g} < 23.4\ \mathrm{mag\,arcsec}^{-2}$ or $R_e < 
2.0''$ are rejected outright; objects in the intermediate range $23.4 \le 
\mu_{\mathrm{eff},g} < 24.3\ \mathrm{mag\,arcsec}^{-2}$ or $2.0'' \le R_e < 2.5''$ 
are retained as a supplementary sample for sensitivity studies but are not included 
in the Gold catalog.

Candidates passing these structural criteria are subsequently classified by their 
integrated $(g-i)$ colour, evaluated in the outer elliptical annulus ($r > 
0.8\,R_e$), into three classes: \emph{blue} ($g-i < 0.5$), \emph{green} ($0.5 \le 
g-i < 0.8$), and \emph{red} ($g-i \ge 0.8$). Extreme colours ($g-i > 1.5$, $g-i < 
-0.25$, $g-r > 1.25$, or $g-r < -0.5$) are treated as evidence of non-physical 
fits or severe contamination and are rejected. Blue galaxies are included 
automatically. For green galaxies, an additional colour-contrast test is applied: a 
colour excess $\Delta(g-i) = (g-i)_\mathrm{center} - (g-i)_\mathrm{total} > 0.10$ 
indicates a red core inconsistent with a spatially uniform diffuse stellar 
population; such objects, together with all red galaxies, are subject to visual 
inspection. Green galaxies without a significant central colour excess are accepted 
automatically.

Visual inspection focuses on two rejection criteria: the presence of spiral 
structure in the \textsl{galfitm} residual image, indicating a rotationally 
supported disk rather than a diffuse system, and an obviously poor single-component 
fit visible as large-scale coherent residuals. Objects exhibiting either feature are 
moved to the rejected sample.

\begin{table}[h]
\centering
\caption{Object counts at each stage of the pipeline.}
\label{tab:pipeline_stages}
\begin{tabular}{lll}
\hline
Stage & Criterion & $N$\\
\hline
Detection      & SEP ($\sigma=0.6$, minarea $=80$)             & $76{,}982{,}201$ \\
cCN filter     & $p \geq 0.95$                                 & $592{,}739$      \\
$\mu$ filter   & $16 < \mu_\mathrm{central} < 30$              & $584{,}846$      \\
Ring filter    & $f_\mathrm{ring}=0.15,\ f_\mathrm{min}=0.05$ & $557{,}106$      \\
GALFITM        & Converged fits                                & $541{,}256$      \\
Gold catalog   & Structural \& colour cuts                     & $5{,}156$        \\
\hline
\end{tabular}
\end{table}

\section{Crossmatch and empirical pipeline performance}
\label{sec:Crossmatch}

To quantify the empirical selection behaviour of the pipeline described in 
Section~\ref{sec::Pipeline}, we perform an object-level crossmatch with the LSBG 
samples of \citet{Greco2018} and \citet{Tanoglidis2021}. The crossmatch is carried 
out in celestial coordinates using a fixed matching radius of $r_{\mathrm{match}} = 
2.0''$, which is small compared to the typical angular extent of the targeted 
diffuse systems but large compared to the astrometric uncertainties of HSC-SSP 
detections.

\subsection{Sample composition and detection completeness}
\label{sec:stats_recovery}

Processing the full HSC-SSP footprint yields a final catalog of 5156 LSBG candidates 
satisfying the selection criteria defined in Section~\ref{sec:gold_catalog}. Both 
literature catalogs lie entirely within the processed footprint: \citet{Greco2018} 
contribute 781 objects and \citet{Tanoglidis2021} contribute 878 objects.

Of the 781 Greco objects, 149 are recovered in our final catalog, yielding a 
detection completeness of $f_{\mathrm{Greco}} = 149/781 = 19.1\%$. Of the 878 
Tanoglidis objects in the footprint, 92 are recovered, corresponding to $f_{\mathrm{Tanoglidis}} = 
92/878 = 10.5\%$. In total, 241 literature-matched objects appear in the final 
catalog. The lower recovery fraction for the Tanoglidis sample is consistent with 
that catalog containing a higher fraction of faint or highly extended sources 
falling below the \textsl{SEP} detection threshold ($\sigma = 0.6$; 
Section~\ref{subsec::SEP}), indicating that incompleteness is controlled primarily 
by detection sensitivity rather than by subsequent filtering stages.

The remaining 4915 objects ($95.3\%$) are not present in either reference catalog. 
By construction, all 5156 objects satisfy the quantitative LSBG selection criteria 
of Section~\ref{sec:gold_catalog}. The 241 literature matches validate that these 
criteria successfully identify objects consistent with established LSBG definitions, 
while the 4915 additional objects represent either previously uncatalogued LSBGs or 
genuine diffuse systems absent from these particular literature samples due to 
differences in survey strategy or selection function.

\subsection{Pipeline sensitivity and morphological selectivity}
\label{sec:pipeline_sensitivity}

To evaluate pipeline performance quantitatively, we track how literature LSBGs 
propagate through successive filtering stages. The key metric is recall,
\begin{equation}
\mathrm{Recall} = \frac{\mathrm{TP}}{\mathrm{TP} + \mathrm{FN}},
\end{equation}
where TP denotes literature LSBGs passing the pipeline and FN denotes literature 
LSBGs rejected by filtering stages.

The cCN morphological filter (Section~\ref{sec:cnn_training}), applied with a 
conservative decision threshold ($p \geq 0.95$), achieves $\mathrm{Recall_{cCN}} = 
1.00$ with respect to all 241 literature LSBGs that pass \textsl{SEP} detection. 
All literature-matched objects, once detected, pass the morphological filter without 
loss. The cCN simultaneously reduces the full detection set by more than four orders 
of magnitude (from ${\sim}77 \times 10^6$ to ${\sim}593{,}000$ candidates), 
demonstrating that the network effectively distinguishes diffuse extended 
morphologies from compact and structured sources despite its aggressive filtering of 
the general detection population.

The subsequent physical pre-filters and \textsl{galfitm} structural modelling 
introduce no additional losses for literature LSBGs, maintaining 
$\mathrm{Recall_{final}} = 1.00$ through to the final catalog. The dominant source 
of incompleteness therefore lies exclusively at the initial \textsl{SEP} detection 
stage. The pipeline is best characterised as detection-limited rather than 
selection-limited: once a diffuse system is detected, it is reliably retained 
through all subsequent stages.

\section{Statistical analysis}
\label{sec:statistics}

This section characterizes the observed-frame photometric and structural properties 
of the Gold catalog and evaluates its consistency with previously published 
wide-area LSBG samples. All quantities are directly observable (colours, surface 
brightness, angular size, and single-component S\'ersic parameters), enabling a 
homogeneous comparison with HSC-SSP and DES-based LSBG catalogs 
\citep{Greco2018, Tanoglidis2021}.

\subsection{Colour distribution and bimodality}
\label{sec:stats_colors}

\begin{figure}
    \centering
    \includegraphics[width=\linewidth]{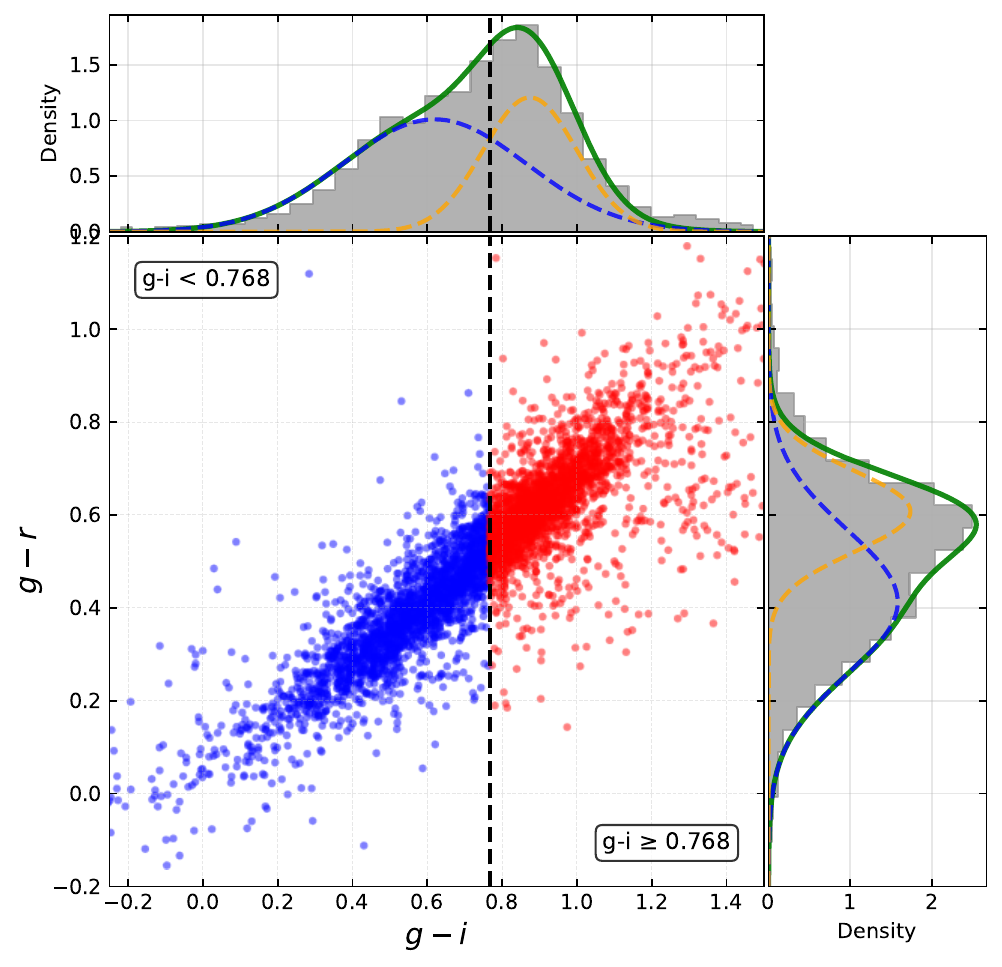}
    \caption{$g-i$ versus $g-r$ distribution of the Gold catalog with 
    one-dimensional marginals. The vertical dashed line marks the adopted 
    separator at $(g-i)_{\rm split} = 0.768$, identified as the minimum between 
    the two colour populations. Blue and red points correspond to the two 
    subsamples defined by this separator.}
    \label{fig:color_color}
\end{figure}

Figure~\ref{fig:color_color} shows the $g-i$ versus $g-r$ distribution of the Gold 
catalog. The objects populate a narrow locus consistent with previously reported 
LSBG samples \citep{Greco2018, Tanoglidis2021}, indicating that the colour 
properties are not dominated by artifacts or extreme outliers.

The one-dimensional $g-i$ distribution exhibits a clear bimodality. We identify the 
minimum of the valley between the two populations at
\begin{equation}
(g-i)_{\rm split} = 0.768,
\end{equation}
which we adopt to define blue and red subsamples for the statistical analysis 
presented in this section. This separator is distinct from the blue/green/red 
classification used in the pipeline construction (Section~\ref{sec:gold_catalog}), 
which serves as a selection criterion; here it reflects the empirical colour 
structure of the resulting catalog. The split yields
\begin{equation}
N_{\rm blue} = 2575, \qquad N_{\rm red} = 2557.
\end{equation}

The existence of a bimodal colour distribution is in agreement with earlier findings 
for LSBGs in both HSC-SSP and DES \citep{Greco2018, Tanoglidis2021}. The exact location 
of the separator is shifted relative to previous studies ($g-i \sim 0.6$--$0.66$), 
which can be attributed to differences in photometric measurements and sample 
construction. Importantly, both populations are clearly present and roughly equal in 
size, indicating that the selection pipeline does not preferentially suppress either 
blue or red systems.

\subsection{Structural parameter space}
\label{sec:stats_structure}

\begin{figure}
    \centering
    \includegraphics[width=\linewidth]{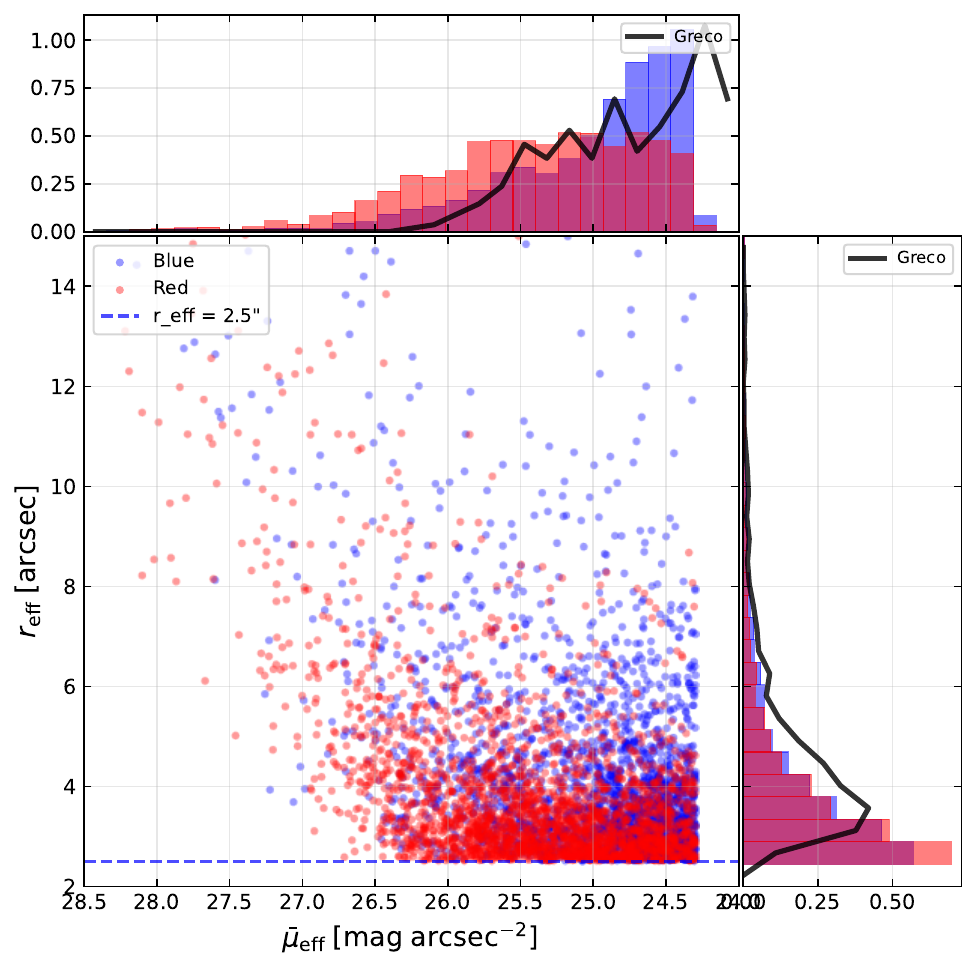}
    \caption{Mean effective surface brightness $\bar{\mu}_{\rm eff}(g)$ versus 
    effective radius $r_{\rm eff}$ for the Gold catalog. Blue and red points 
    correspond to the colour subsamples defined in 
    Section~\ref{sec:stats_colors}. The dashed horizontal line marks the 
    adopted lower limit $r_{\rm eff} = 2.5''$. The black curve in the marginal 
    histograms shows the \citet{Greco2018} comparison sample.}
    \label{fig:mueffAndre}
\end{figure}

The joint distribution of mean effective surface brightness $\bar{\mu}_{\rm eff}(g)$ 
and effective radius $r_{\rm eff}$ is shown in Fig.~\ref{fig:mueffAndre}. The 
catalog spans a broad range in angular size ($r_{\rm eff} \simeq 2.5''$ to 
${\sim}14''$) and extends to very low surface brightness values 
($\bar{\mu}_{\rm eff} \gtrsim 29\ \mathrm{mag\,arcsec^{-2}}$), fully covering the 
regime associated with diffuse galaxies in previous surveys. The distribution is 
continuous and does not show a dominant accumulation at the imposed selection 
boundaries, suggesting that the sample reflects an underlying population with a 
smooth distribution in structural parameters rather than one strongly truncated by 
the applied cuts. Both blue and red subsamples occupy largely overlapping regions 
of this parameter space, with no strong colour-dependent stratification in size or 
surface brightness.

Comparison with the \citet{Greco2018} sample (shown as the black curve in the 
marginal histograms) reveals broad agreement in the $r_{\rm eff}$ distribution, 
while our catalog extends to fainter surface brightnesses, consistent with the 
deeper effective detection threshold enabled by the permissive \textsl{SEP} 
detection combined with cCN filtering.

\begin{figure}
    \centering
    \includegraphics[width=\linewidth]{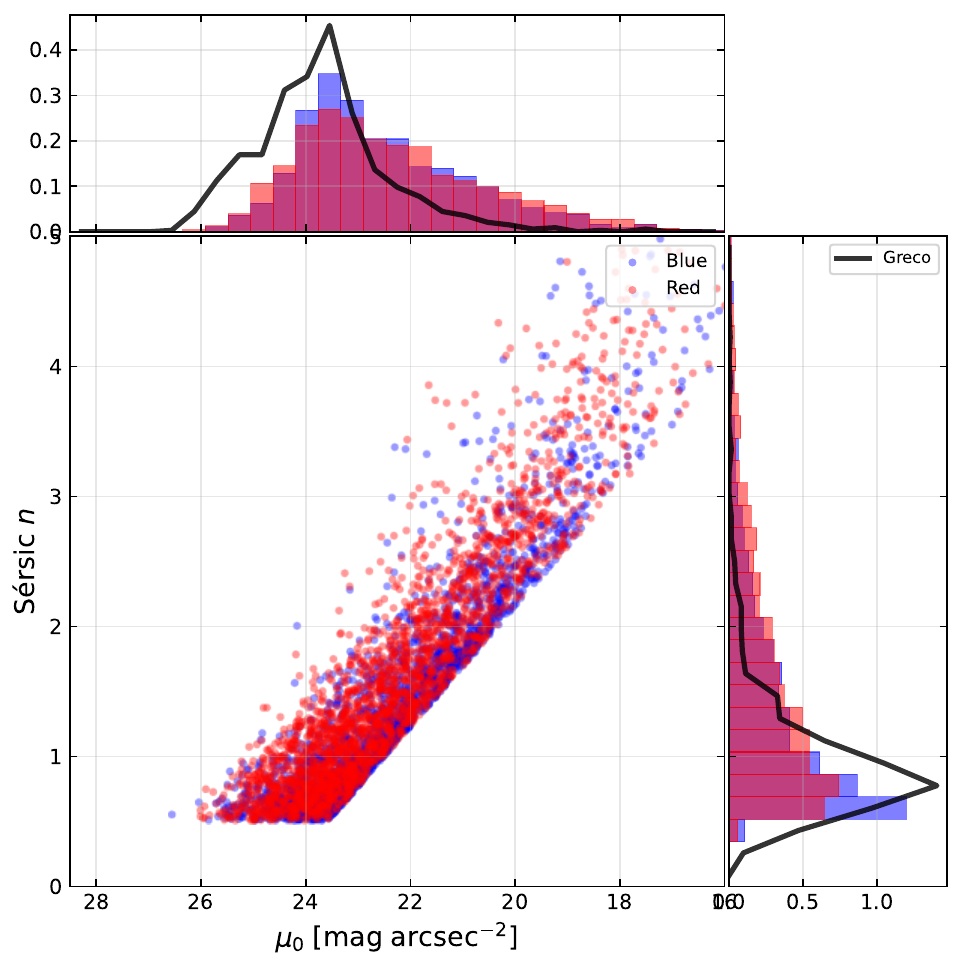}
    \caption{Central surface brightness $\mu_0(g)$ versus S\'ersic index $n$ 
    for the Gold catalog. Blue and red points correspond to the colour 
    subsamples defined in Section~\ref{sec:stats_colors}. The black curve in 
    the marginal histograms shows the \citet{Greco2018} comparison sample.}
    \label{fig:mu0Andn}
\end{figure}

In the $\mu_0(g)$ versus $n$ plane (Fig.~\ref{fig:mu0Andn}), most objects exhibit 
low S\'ersic indices ($n \sim 0.5$--$3$), consistent with exponential or 
near-exponential profiles characteristic of diffuse disk-like systems. A mild trend 
is visible in which higher central surface brightness systems extend toward larger 
$n$, while the lowest surface-brightness objects are predominantly found at low 
S\'ersic indices. This behaviour is consistent with results from HSC-SSP and DES LSBG 
samples \citep{Greco2018, Tanoglidis2021} and argues against significant 
contamination by compact background galaxies or poorly fitted objects. The 
\citet{Greco2018} comparison sample peaks at $n \sim 0.7$--$0.8$, in good agreement 
with our distribution.

\subsection{Spectroscopic validation and redshift distribution}
\label{sec:stats_specz}

To assess the redshift properties of the final catalog, we cross-matched our LSBG candidates with spectroscopic databases from DESI (\cite{DESI1},\cite{DESI2}). Of the 5156 final candidates, 248 ($4.8\%$) have spectroscopic redshift measurements available. These spectroscopically confirmed objects exhibit a median redshift of $\bar{z}_{\rm spec} = 0.122$ and a median effective surface brightness of $\bar{\mu}_{\rm eff} = 24.61\ \mathrm{mag\,arcsec}^{-2}$, placing them at the bright end of the catalog.

To quantify the selection bias inherent in this subsample, we compare its redshift distribution with that of a random sample of 20{,}000 DESI spectroscopic objects, which has a median redshift of $\bar{z}_{\rm DESI} = 0.506$. The substantial offset ($\Delta\bar{z} \approx 0.38$) confirms that spectroscopic follow-up has preferentially targeted nearby, brighter systems, and that the 248 confirmed objects are not representative of the full catalog. The remaining 4908 objects ($95.2\%$) lack spectroscopic confirmation and likely include a higher fraction of more distant or intrinsically fainter sources. Because of the size of the objects it is more plausible to expect a higher fraction of fainter sources than a high fraction of distant objects.

The spectroscopic subsample nonetheless provides direct validation that the pipeline identifies genuine low-surface-brightness galaxies in the local universe. Future spectroscopic observations of a representative subset would be valuable for characterising the true redshift distribution and luminosity properties of the full population.

\section{Conclusions}
\label{sec:conclusions}

We have presented a hybrid detection pipeline for low-surface-brightness galaxies in HSC-SSP imaging that integrates a cCN as a morphological validation stage alongside permissive low-threshold source detection, physically motivated pre-filters, and parametric surface-brightness modelling with \textsl{galfitm}. The central design choice is to deliberately constrain the classifier to ${\sim}\,10^4$ trainable parameters, matching its representational capacity to the spatial scales and signal characteristics of diffuse emission rather than employing deep general-purpose architectures.

The primary goal of this work was to demonstrate that a parameter-efficient cCN can serve as an effective morphological filter within a realistic survey pipeline, and this goal is met convincingly. Applied with a conservative decision threshold of $p \geq 0.95$, the cCN reduces the initial \textsl{SEP} detection set of ${\sim}77 \times 10^6$ objects by more than four orders of magnitude while achieving a recall of $\mathrm{Recall_{cCN}} = 1.00$ with respect to all literature LSBGs passing the detection stage. All detected literature-matched objects are retained through every subsequent filtering stage, confirming that the cCN suppresses contaminants without sacrificing sensitivity to genuine diffuse systems. The resulting Gold catalog of 5156 candidates occupies the characteristic loci of LSBGs in colour--colour and structural parameter space, with a bimodal $g-i$ distribution and predominantly low S\'ersic indices consistent with near-exponential profiles, in close agreement with previously published HSC-SSP-based samples \citep{Greco2018, Tanoglidis2021}.

The overall pipeline recovery fraction of $C_{\rm total} \approx 0.10$ with respect 
to the \citet{Tanoglidis2021} reference sample demonstrates that incompleteness 
remains substantial. However, as shown in Section~\ref{sec:Crossmatch}, this 
incompleteness arises exclusively at the initial \textsl{SEP} detection stage. The 
pipeline is therefore best characterised as detection-limited rather than 
selection-limited: once a diffuse system is detected, it is reliably carried through 
all subsequent stages.

Several limitations constrain the interpretation of these results. The positive 
training sample is drawn exclusively from the \citet{Greco2018} catalogue, meaning 
the classifier is optimised to recover objects whose morphology resembles that prior; 
diffuse systems occupying different regions of parameter space may be systematically 
suppressed. The test-set performance therefore measures interpolation within the 
training distribution rather than generalisation to an independent LSBG population. 
Furthermore, the reported purity estimates should be understood as lower bounds that 
depend on the completeness of the reference catalogues used for evaluation.

Notwithstanding these caveats, the results support the conclusion that 
parameter-efficient cCNs provide a practical and computationally scalable 
morphological filter within multi-stage LSBG pipelines. In this sense, cCNs occupy 
a useful middle ground between purely classical pipelines and large, general-purpose 
deep architectures: expressive enough to suppress the dominant failure modes of 
permissive detection, yet compact enough to remain robust when deployed at the scale 
of wide-area imaging surveys.

Immediate priorities for follow-up work include: (i) systematic improvement of the \textsl{galfitm} branch through more robust initialisation strategies and improved neighbour treatment, which currently accounts for a significant fraction of candidate losses; (ii) extension of the machine-learning component to ensemble or multi-head configurations trained on complementary sample definitions, in order to reduce sensitivity to the morphological prior imposed by the single catalogue training set; and (iii) the long-term development of a fully compact network driven detection stage that operates directly on survey imaging without prior source extraction, removing the current dependence on \textsl{SEP} and the associated detection stage incompleteness that represents the dominant limitation of the present pipeline.

\begin{acknowledgements}\\
The Hyper Suprime-Cam (HSC) collaboration includes the astronomical communities of Japan and Taiwan, and Princeton University. The HSC instrumentation and software were developed by the National Astronomical Observatory of Japan (NAOJ), the Kavli Institute for the Physics and Mathematics of the Universe (Kavli IPMU), the University of Tokyo, the High Energy Accelerator Research Organization (KEK), the Academia Sinica Institute for Astronomy and Astrophysics in Taiwan (ASIAA), and Princeton University. Funding was contributed by the FIRST program from the Japanese Cabinet Office, the Ministry of Education, Culture, Sports, Science and Technology (MEXT), the Japan Society for the Promotion of Science (JSPS), Japan Science and Technology Agency (JST), the Toray Science Foundation, NAOJ, Kavli IPMU, KEK, ASIAA, and Princeton University.\\
\\
This paper makes use of software developed for Vera C. Rubin Observatory. We thank the Rubin Observatory for making their code available as free software at http://pipelines.lsst.io/.\\
This paper is based on data collected at the Subaru Telescope and retrieved from the HSC data archive system, which is operated by the Subaru Telescope and Astronomy Data Center (ADC) at NAOJ. Data analysis was in part carried out with the cooperation of Center for Computational Astrophysics (CfCA), NAOJ. We are honored and grateful for the opportunity of observing the Universe from Maunakea, which has the cultural, historical and natural significance in Hawaii. \\
\\
This research used data obtained with the Dark Energy Spectroscopic Instrument (DESI). DESI construction and operations is managed by the Lawrence Berkeley National Laboratory. This material is based upon work supported by the U.S. Department of Energy, Office of Science, Office of High-Energy Physics, under Contract No. DE–AC02–05CH11231, and by the National Energy Research Scientific Computing Center, a DOE Office of Science User Facility under the same contract. Additional support for DESI was provided by the U.S. National Science Foundation (NSF), Division of Astronomical Sciences under Contract No. AST-0950945 to the NSF’s National Optical-Infrared Astronomy Research Laboratory; the Science and Technology Facilities Council of the United Kingdom; the Gordon and Betty Moore Foundation; the Heising-Simons Foundation; the French Alternative Energies and Atomic Energy Commission (CEA); the National Council of Humanities, Science and Technology of Mexico (CONAHCYT); the Ministry of Science and Innovation of Spain (MICINN), and by the DESI Member Institutions: www.desi.lbl.gov/collaborating-institutions. The DESI collaboration is honored to be permitted to conduct scientific research on I’oligam Du’ag (Kitt Peak), a mountain with particular significance to the Tohono O’odham Nation. Any opinions, findings, and conclusions or recommendations expressed in this material are those of the author(s) and do not necessarily reflect the views of the U.S. National Science Foundation, the U.S. Department of Energy, or any of the listed funding agencies.

\end{acknowledgements}

%
%

\bibliographystyle{aa}
\bibliography{references}

@ARTICLE{DES,
       author = {{Abbott}, T.~M.~C. and {Adam{\'o}w}, M. and {Aguena}, M. and {Allam}, S. and {Amon}, A. and {Annis}, J. and {Avila}, S. and {Bacon}, D. and {Banerji}, M. and {Bechtol}, K. and {Becker}, M.~R. and {Bernstein}, G.~M. and {Bertin}, E. and {Bhargava}, S. and {Bridle}, S.~L. and {Brooks}, D. and {Burke}, D.~L. and {Carnero Rosell}, A. and {Carrasco Kind}, M. and {Carretero}, J. and {Castander}, F.~J. and {Cawthon}, R. and {Chang}, C. and {Choi}, A. and {Conselice}, C. and {Costanzi}, M. and {Crocce}, M. and {da Costa}, L.~N. and {Davis}, T.~M. and {De Vicente}, J. and {DeRose}, J. and {Desai}, S. and {Diehl}, H.~T. and {Dietrich}, J.~P. and {Drlica-Wagner}, A. and {Eckert}, K. and {Elvin-Poole}, J. and {Everett}, S. and {Evrard}, A.~E. and {Ferrero}, I. and {Fert{\'e}}, A. and {Flaugher}, B. and {Fosalba}, P. and {Friedel}, D. and {Frieman}, J. and {Garc{\'\i}a-Bellido}, J. and {Gaztanaga}, E. and {Gelman}, L. and {Gerdes}, D.~W. and {Giannantonio}, T. and {Gill}, M.~S.~S. and {Gruen}, D. and {Gruendl}, R.~A. and {Gschwend}, J. and {Gutierrez}, G. and {Hartley}, W.~G. and {Hinton}, S.~R. and {Hollowood}, D.~L. and {Honscheid}, K. and {Huterer}, D. and {James}, D.~J. and {Jeltema}, T. and {Johnson}, M.~D. and {Kent}, S. and {Kron}, R. and {Kuehn}, K. and {Kuropatkin}, N. and {Lahav}, O. and {Li}, T.~S. and {Lidman}, C. and {Lin}, H. and {MacCrann}, N. and {Maia}, M.~A.~G. and {Manning}, T.~A. and {Maloney}, J.~D. and {March}, M. and {Marshall}, J.~L. and {Martini}, P. and {Melchior}, P. and {Menanteau}, F. and {Miquel}, R. and {Morgan}, R. and {Myles}, J. and {Neilsen}, E. and {Ogando}, R.~L.~C. and {Palmese}, A. and {Paz-Chinch{\'o}n}, F. and {Petravick}, D. and {Pieres}, A. and {Plazas}, A.~A. and {Pond}, C. and {Rodriguez-Monroy}, M. and {Romer}, A.~K. and {Roodman}, A. and {Rykoff}, E.~S. and {Sako}, M. and {Sanchez}, E. and {Santiago}, B. and {Scarpine}, V. and {Serrano}, S. and {Sevilla-Noarbe}, I. and {Smith}, J. Allyn and {Smith}, M. and {Soares-Santos}, M. and {Suchyta}, E. and {Swanson}, M.~E.~C. and {Tarle}, G. and {Thomas}, D. and {To}, C. and {Tremblay}, P.~E. and {Troxel}, M.~A. and {Tucker}, D.~L. and {Turner}, D.~J. and {Varga}, T.~N. and {Walker}, A.~R. and {Wechsler}, R.~H. and {Weller}, J. and {Wester}, W. and {Wilkinson}, R.~D. and {Yanny}, B. and {Zhang}, Y. and {Nikutta}, R. and {Fitzpatrick}, M. and {Jacques}, A. and {Scott}, A. and {Olsen}, K. and {Huang}, L. and {Herrera}, D. and {Juneau}, S. and {Nidever}, D. and {Weaver}, B.~A. and {Adean}, C. and {Correia}, V. and {de Freitas}, M. and {Freitas}, F.~N. and {Singulani}, C. and {Vila-Verde}, G. and {Linea Science Server}},
        title = "{The Dark Energy Survey Data Release 2}",
      journal = {\apjs},
         year = 2021,
        month = aug,
       volume = {255},
       number = {2},
          eid = {20},
        pages = {20},
          doi = {10.3847/1538-4365/ac00b3},
archivePrefix = {arXiv},
       eprint = {2101.05765},
 primaryClass = {astro-ph.IM},
       adsurl = {https://ui.adsabs.harvard.edu/abs/2021ApJS..255...20A}
}

@article{Aihara2018,
    author = {Aihara, Hiroaki and Arimoto, Nobuo and Armstrong, Robert and Arnouts, Stéphane and Bahcall, Neta A and Bickerton, Steven and Bosch, James and Bundy, Kevin and Capak, Peter L and Chan, James H H and Chiba, Masashi and Coupon, Jean and Egami, Eiichi and Enoki, Motohiro and Finet, Francois and Fujimori, Hiroki and Fujimoto, Seiji and Furusawa, Hisanori and Furusawa, Junko and Goto, Tomotsugu and Goulding, Andy and Greco, Johnny P and Greene, Jenny E and Gunn, James E and Hamana, Takashi and Harikane, Yuichi and Hashimoto, Yasuhiro and Hattori, Takashi and Hayashi, Masao and Hayashi, Yusuke and Hełminiak, Krzysztof G and Higuchi, Ryo and Hikage, Chiaki and Ho, Paul T P and Hsieh, Bau-Ching and Huang, Kuiyun and Huang, Song and Ikeda, Hiroyuki and Imanishi, Masatoshi and Inoue, Akio K and Iwasawa, Kazushi and Iwata, Ikuru and Jaelani, Anton T and Jian, Hung-Yu and Kamata, Yukiko and Karoji, Hiroshi and Kashikawa, Nobunari and Katayama, Nobuhiko and Kawanomoto, Satoshi and Kayo, Issha and Koda, Jin and Koike, Michitaro and Kojima, Takashi and Komiyama, Yutaka and Konno, Akira and Koshida, Shintaro and Koyama, Yusei and Kusakabe, Haruka and Leauthaud, Alexie and Lee, Chien-Hsiu and Lin, Lihwai and Lin, Yen-Ting and Lupton, Robert H and Mandelbaum, Rachel and Matsuoka, Yoshiki and Medezinski, Elinor and Mineo, Sogo and Miyama, Shoken and Miyatake, Hironao and Miyazaki, Satoshi and Momose, Rieko and More, Anupreeta and More, Surhud and Moritani, Yuki and Moriya, Takashi J and Morokuma, Tomoki and Mukae, Shiro and Murata, Ryoma and Murayama, Hitoshi and Nagao, Tohru and Nakata, Fumiaki and Niida, Mana and Niikura, Hiroko and Nishizawa, Atsushi J and Obuchi, Yoshiyuki and Oguri, Masamune and Oishi, Yukie and Okabe, Nobuhiro and Okamoto, Sakurako and Okura, Yuki and Ono, Yoshiaki and Onodera, Masato and Onoue, Masafusa and Osato, Ken and Ouchi, Masami and Price, Paul A and Pyo, Tae-Soo and Sako, Masao and Sawicki, Marcin and Shibuya, Takatoshi and Shimasaku, Kazuhiro and Shimono, Atsushi and Shirasaki, Masato and Silverman, John D and Simet, Melanie and Speagle, Joshua and Spergel, David N and Strauss, Michael A and Sugahara, Yuma and Sugiyama, Naoshi and Suto, Yasushi and Suyu, Sherry H and Suzuki, Nao and Tait, Philip J and Takada, Masahiro and Takata, Tadafumi and Tamura, Naoyuki and Tanaka, Manobu M and Tanaka, Masaomi and Tanaka, Masayuki and Tanaka, Yoko and Terai, Tsuyoshi and Terashima, Yuichi and Toba, Yoshiki and Tominaga, Nozomu and Toshikawa, Jun and Turner, Edwin L and Uchida, Tomohisa and Uchiyama, Hisakazu and Umetsu, Keiichi and Uraguchi, Fumihiro and Urata, Yuji and Usuda, Tomonori and Utsumi, Yousuke and Wang, Shiang-Yu and Wang, Wei-Hao and Wong, Kenneth C and Yabe, Kiyoto and Yamada, Yoshihiko and Yamanoi, Hitomi and Yasuda, Naoki and Yeh, Sherry and Yonehara, Atsunori and Yuma, Suraphong},
    title = {The Hyper Suprime-Cam SSP Survey: Overview and survey design},
    journal = {Publications of the Astronomical Society of Japan},
    volume = {70},
    number = {SP1},
    pages = {S4},
    year = {2018},
    month = {01},
    issn = {0004-6264},
    doi = {10.1093/pasj/psx066},
    url = {https://doi.org/10.1093/pasj/psx066},
    eprint = {https://academic.oup.com/pasj/article-pdf/70/SP1/S4/54675128/pasj_70_sp1_s4.pdf},
}

@article{Aihara2022,
    author = {Aihara, Hiroaki and AlSayyad, Yusra and Ando, Makoto and Armstrong, Robert and Bosch, James and Egami, Eiichi and Furusawa, Hisanori and Furusawa, Junko and Harasawa, Sumiko and Harikane, Yuichi and Hsieh, Bau-Ching and Ikeda, Hiroyuki and Ito, Kei and Iwata, Ikuru and Kodama, Tadayuki and Koike, Michitaro and Kokubo, Mitsuru and Komiyama, Yutaka and Li, Xiangchong and Liang, Yongming and Lin, Yen-Ting and Lupton, Robert H and Lust, Nate B and MacArthur, Lauren A and Mawatari, Ken and Mineo, Sogo and Miyatake, Hironao and Miyazaki, Satoshi and More, Surhud and Morishima, Takahiro and Murayama, Hitoshi and Nakajima, Kimihiko and Nakata, Fumiaki and Nishizawa, Atsushi J and Oguri, Masamune and Okabe, Nobuhiro and Okura, Yuki and Ono, Yoshiaki and Osato, Ken and Ouchi, Masami and Pan, Yen-Chen and Plazas Malagón, Andrés A and Price, Paul A and Reed, Sophie L and Rykoff, Eli S and Shibuya, Takatoshi and Simunovic, Mirko and Strauss, Michael A and Sugimori, Kanako and Suto, Yasushi and Suzuki, Nao and Takada, Masahiro and Takagi, Yuhei and Takata, Tadafumi and Takita, Satoshi and Tanaka, Masayuki and Tang, Shenli and Taranu, Dan S and Terai, Tsuyoshi and Toba, Yoshiki and Turner, Edwin L and Uchiyama, Hisakazu and Vijarnwannaluk, Bovornpratch and Waters, Christopher Z and Yamada, Yoshihiko and Yamamoto, Naoaki and Yamashita, Takuji},
    title = {Third data release of the Hyper Suprime-Cam Subaru Strategic Program},
    journal = {Publications of the Astronomical Society of Japan},
    volume = {74},
    number = {2},
    pages = {247-272},
    year = {2022},
    month = {04},
    issn = {0004-6264},
    doi = {10.1093/pasj/psab122},
    url = {https://doi.org/10.1093/pasj/psab122},
    eprint = {https://academic.oup.com/pasj/article-pdf/74/2/247/54647221/psab122.pdf},
}

@article{Akhlaghi2015,
doi = {10.1088/0067-0049/220/1/1},
url = {https://doi.org/10.1088/0067-0049/220/1/1},
year = {2015},
month = {aug},
publisher = {The American Astronomical Society},
volume = {220},
number = {1},
pages = {1},
author = {Akhlaghi, Mohammad and Ichikawa, Takashi},
title = {NOISE-BASED DETECTION AND SEGMENTATION OF NEBULOUS OBJECTS},
journal = {The Astrophysical Journal Supplement Series}
}

@software{Akiba2019Optuna:Framework,
       author = {{Akiba}, Takuya and {Sano}, Shotaro and {Yanase}, Toshihiko and {Ohta}, Takeru and {Koyama}, Masanori},
        title = "{Optuna: A Next-generation Hyperparameter Optimization Framework}",
      journal = {arXiv e-prints},
         year = 2019,
        month = jul,
          eid = {arXiv:1907.10902},
        pages = {arXiv:1907.10902},
          doi = {10.48550/arXiv.1907.10902},
archivePrefix = {arXiv},
       eprint = {1907.10902},
 primaryClass = {cs.LG},
       adsurl = {https://ui.adsabs.harvard.edu/abs/2019arXiv190710902A}
}

@article{Barbary2016, doi = {10.21105/joss.00058}, url = {https://doi.org/10.21105/joss.00058}, year = {2016}, publisher = {The Open Journal}, volume = {1}, number = {6}, pages = {58}, author = {Barbary, Kyle}, title = {SEP: Source Extractor as a library}, journal = {Journal of Open Source Software} }

@article{ Bertin1996,
	author = {{Bertin, E.} and {Arnouts, S.}},
	title = {SExtractor: Software for source extraction},
	DOI= "10.1051/aas:1996164",
	url= "https://doi.org/10.1051/aas:1996164",
	journal = {Astron. Astrophys. Suppl. Ser.},
	year = 1996,
	volume = 117,
	number = 2,
	pages = "393-404",
}

@article{Blanton2005,
doi = {10.1086/431416},
url = {https://doi.org/10.1086/431416},
year = {2005},
month = {sep},
publisher = {},
volume = {631},
number = {1},
pages = {208},
author = {Blanton, Michael R. and Lupton, Robert H. and Schlegel, David J. and Strauss, Michael A. and Brinkmann, J. and Fukugita, Masataka and Loveday, Jon},
title = {The Properties and Luminosity Function of Extremely Low Luminosity Galaxies*},
journal = {The Astrophysical Journal}
}

@ARTICLE{Bothun1987,
       author = {{Bothun}, Gregory D. and {Impey}, Christopher D. and {Malin}, David F. and {Mould}, Jeremy R.},
        title = "{Discovery of a Huge Low-Surface-Brightness Galaxy: A Proto-Disk Galaxy at Low Redshift?}",
      journal = {\aj},
         year = 1987,
        month = jul,
       volume = {94},
        pages = {23},
          doi = {10.1086/114443},
       adsurl = {https://ui.adsabs.harvard.edu/abs/1987AJ.....94...23B}
}

@article{Boylan-Kolchin2011,
    author = {Boylan-Kolchin, Michael and Bullock, James S. and Kaplinghat, Manoj},
    title = {Too big to fail? The puzzling darkness of massive Milky Way subhaloes},
    journal = {Monthly Notices of the Royal Astronomical Society: Letters},
    volume = {415},
    number = {1},
    pages = {L40-L44},
    year = {2011},
    month = {07},
    issn = {1745-3925},
    doi = {10.1111/j.1745-3933.2011.01074.x},
    url = {https://doi.org/10.1111/j.1745-3933.2011.01074.x},
    eprint = {https://academic.oup.com/mnrasl/article-pdf/415/1/L40/54669362/mnrasl_415_1_l40.pdf},
}

@article{Bullock2017,
   author = "Bullock, James S. and Boylan-Kolchin, Michael",
   title = "Small-Scale Challenges to the ΛCDM Paradigm", 
   journal= "Annual Review of Astronomy and Astrophysics",
   year = "2017",
   volume = "55",
   number = "Volume 55, 2017",
   pages = "343-387",
   doi = "https://doi.org/10.1146/annurev-astro-091916-055313",
   url = "https://www.annualreviews.org/content/journals/10.1146/annurev-astro-091916-055313",
   publisher = "Annual Reviews",
   issn = "1545-4282",
   type = "Journal Article",
  }

@inproceedings{Chollet2017,
  author    = {Chollet, F.},
  title     = {Xception: Deep Learning with Depthwise Separable Convolutions},
  booktitle = {CVPR},
  year      = {2017},
  pages     = {1251--1258},
  doi       = {10.1109/CVPR.2017.195}
}

@ARTICLE{DalCanton1997,
       author = {{Dalcanton}, Julianne J. and {Spergel}, David N. and {Summers}, F.~J.},
        title = "{The Formation of Disk Galaxies}",
      journal = {\apj},
         year = 1997,
        month = jun,
       volume = {482},
       number = {2},
        pages = {659-676},
          doi = {10.1086/304182},
archivePrefix = {arXiv},
       eprint = {astro-ph/9611226},
 primaryClass = {astro-ph},
       adsurl = {https://ui.adsabs.harvard.edu/abs/1997ApJ...482..659D}
}

@ARTICLE{DESI1,
       author = {{DESI Collaboration} and {Adame}, A.~G. and {Aguilar}, J. and {Ahlen}, S. and {Alam}, S. and {Aldering}, G. and {Alexander}, D.~M. and {Alfarsy}, R. and {Allende Prieto}, C. and {Alvarez}, M. and {Alves}, O. and {Anand}, A. and {Andrade-Oliveira}, F. and {Armengaud}, E. and {Asorey}, J. and {Avila}, S. and {Aviles}, A. and {Bailey}, S. and {Balaguera-Antol{\'\i}nez}, A. and {Ballester}, O. and {Baltay}, C. and {Bault}, A. and {Bautista}, J. and {Behera}, J. and {Beltran}, S.~F. and {BenZvi}, S. and {Beraldo e Silva}, L. and {Bermejo-Climent}, J.~R. and {Berti}, A. and {Besuner}, R. and {Beutler}, F. and {Bianchi}, D. and {Blake}, C. and {Blum}, R. and {Bolton}, A.~S. and {Brieden}, S. and {Brodzeller}, A. and {Brooks}, D. and {Brown}, Z. and {Buckley-Geer}, E. and {Burtin}, E. and {Cabayol-Garcia}, L. and {Cai}, Z. and {Canning}, R. and {Cardiel-Sas}, L. and {Carnero Rosell}, A. and {Castander}, F.~J. and {Cervantes-Cota}, J.~L. and {Chabanier}, S. and {Chaussidon}, E. and {Chaves-Montero}, J. and {Chen}, S. and {Chen}, X. and {Chuang}, C. and {Claybaugh}, T. and {Cole}, S. and {Cooper}, A.~P. and {Cuceu}, A. and {Davis}, T.~M. and {Dawson}, K. and {de Belsunce}, R. and {de la Cruz}, R. and {de la Macorra}, A. and {Della Costa}, J. and {de Mattia}, A. and {Demina}, R. and {Demirbozan}, U. and {DeRose}, J. and {Dey}, A. and {Dey}, B. and {Dhungana}, G. and {Ding}, J. and {Ding}, Z. and {Doel}, P. and {Doshi}, R. and {Douglass}, K. and {Edge}, A. and {Eftekharzadeh}, S. and {Eisenstein}, D.~J. and {Elliott}, A. and {Ereza}, J. and {Escoffier}, S. and {Fagrelius}, P. and {Fan}, X. and {Fanning}, K. and {Fawcett}, V.~A. and {Ferraro}, S. and {Flaugher}, B. and {Font-Ribera}, A. and {Forero-Romero}, J.~E. and {Forero-S{\'a}nchez}, D. and {Frenk}, C.~S. and {G{\"a}nsicke}, B.~T. and {Garc{\'\i}a}, L. {\'A}. and {Garc{\'\i}a-Bellido}, J. and {Garcia-Quintero}, C. and {Garrison}, L.~H. and {Gil-Mar{\'\i}n}, H. and {Golden-Marx}, J. and {Gontcho A Gontcho}, S. and {Gonzalez-Morales}, A.~X. and {Gonzalez-Perez}, V. and {Gordon}, C. and {Graur}, O. and {Green}, D. and {Gruen}, D. and {Guy}, J. and {Hadzhiyska}, B. and {Hahn}, C. and {Han}, J.~J. and {Hanif}, M.~M.~S. and {Herrera-Alcantar}, H.~K. and {Honscheid}, K. and {Hou}, J. and {Howlett}, C. and {Huterer}, D. and {Ir{\v{s}}i{\v{c}}}, V. and {Ishak}, M. and {Jacques}, A. and {Jana}, A. and {Jiang}, L. and {Jimenez}, J. and {Jing}, Y.~P. and {Joudaki}, S. and {Joyce}, R. and {Jullo}, E. and {Juneau}, S. and {Kara{\c{c}}ayl{\i}}, N.~G. and {Karim}, T. and {Kehoe}, R. and {Kent}, S. and {Khederlarian}, A. and {Kim}, S. and {Kirkby}, D. and {Kisner}, T. and {Kitaura}, F. and {Kizhuprakkat}, N. and {Kneib}, J. and {Koposov}, S.~E. and {Kov{\'a}cs}, A. and {Kremin}, A. and {Krolewski}, A. and {L'Huillier}, B. and {Lahav}, O. and {Lambert}, A. and {Lamman}, C. and {Lan}, T.-W. and {Landriau}, M. and {Lang}, D. and {Lange}, J.~U. and {Lasker}, J. and {Leauthaud}, A. and {Le Guillou}, L. and {Levi}, M.~E. and {Li}, T.~S. and {Linder}, E. and {Lyons}, A. and {Magneville}, C. and {Manera}, M. and {Manser}, C.~J. and {Margala}, D. and {Martini}, P. and {McDonald}, P. and {Medina}, G.~E. and {Medina-Varela}, L. and {Meisner}, A. and {Mena-Fern{\'a}ndez}, J. and {Meneses-Rizo}, J. and {Mezcua}, M. and {Miquel}, R. and {Montero-Camacho}, P. and {Moon}, J. and {Moore}, S. and {Moustakas}, J. and {Mueller}, E. and {Mundet}, J. and {Mu{\~n}oz-Guti{\'e}rrez}, A. and {Myers}, A.~D. and {Nadathur}, S. and {Napolitano}, L. and {Neveux}, R. and {Newman}, J.~A. and {Nie}, J. and {Nikutta}, R. and {Niz}, G. and {Norberg}, P. and {Noriega}, H.~E. and {Paillas}, E. and {Palanque-Delabrouille}, N. and {Palmese}, A. and {Pan}, Z. and {Parkinson}, D. and {Penmetsa}, S. and {Percival}, W.~J. and {P{\'e}rez-Fern{\'a}ndez}, A. and {P{\'e}rez-R{\`a}fols}, I. and {Pieri}, M. and {Poppett}, C. and {Porredon}, A. and {Pothier}, S.},
        title = "{The Early Data Release of the Dark Energy Spectroscopic Instrument}",
      journal = {\aj},
         year = 2024,
        month = aug,
       volume = {168},
       number = {2},
          eid = {58},
        pages = {58},
          doi = {10.3847/1538-3881/ad3217},
archivePrefix = {arXiv},
       eprint = {2306.06308},
 primaryClass = {astro-ph.CO},
       adsurl = {https://ui.adsabs.harvard.edu/abs/2024AJ....168...58D}
}

@ARTICLE{DESI2,
       author = {{DESI Collaboration} and {Abdul Karim}, M. and {Adame}, A.~G. and {Aguado}, D. and {Aguilar}, J. and {Ahlen}, S. and {Alam}, S. and {Aldering}, G. and {Alexander}, D.~M. and {Alfarsy}, R. and {Allen}, L. and {Allende Prieto}, C. and {Alves}, O. and {Anand}, A. and {Andrade}, U. and {Armengaud}, E. and {Avila}, S. and {Aviles}, A. and {Awan}, H. and {Bailey}, S. and {Baleato Lizancos}, A. and {Ballester}, O. and {Bault}, A. and {Bautista}, J. and {Bean}, R. and {Behera}, J. and {BenZvi}, S. and {Beraldo e Silva}, L. and {Bermejo-Climent}, J.~R. and {Beutler}, F. and {Bianchi}, D. and {Blake}, C. and {Blum}, R. and {Bolton}, A.~S. and {Bonici}, M. and {Brieden}, S. and {Brodzeller}, A. and {Brooks}, D. and {Buckley-Geer}, E. and {Burtin}, E. and {Bystr{\"o}m}, A. and {Canning}, R. and {Carnero Rosell}, A. and {Carr}, A. and {Carrilho}, P. and {Casas}, L. and {Castander}, F.~J. and {Cereskaite}, R. and {Cervantes-Cota}, J.~L. and {Chaussidon}, E. and {Chaves-Montero}, J. and {Chen}, S. and {Chen}, X. and {Circosta}, C. and {Claybaugh}, T. and {Cole}, S. and {Cooper}, A.~P. and {Cousinou}, M.-C. and {Cuceu}, A. and {Davis}, T.~M. and {Dawson}, K.~S. and {de Belsunce}, R. and {de la Cruz}, R. and {de la Macorra}, A. and {de Mattia}, A. and {Deiosso}, N. and {Della Costa}, J. and {Demina}, R. and {Demirbozan}, U. and {DeRose}, J. and {Dey}, A. and {Dey}, B. and {Ding}, J. and {Ding}, Z. and {Doel}, P. and {Douglass}, K. and {Dowicz}, M. and {Ebina}, H. and {Edelstein}, J. and {Eisenstein}, D.~J. and {Elbers}, W. and {Emas}, N. and {Escoffier}, S. and {Fagrelius}, P. and {Fan}, X. and {Fanning}, K. and {Favole}, G. and {Fawcett}, V.~A. and {Fern{\'a}ndez-Garc{\'\i}a}, E. and {Ferraro}, S. and {Findlay}, N. and {Font-Ribera}, A. and {Forero-Romero}, J.~E. and {Forero-S{\'a}nchez}, D. and {Frenk}, C.~S. and {G{\"a}nsicke}, B.~T. and {Galbany}, L. and {Garc{\'\i}a-Bellido}, J. and {Garcia-Quintero}, C. and {Garrison}, L.~H. and {Gazta{\~n}aga}, E. and {Gil-Mar{\'\i}n}, H. and {Gloudemans}, A. and {Gnedin}, O.~Y. and {Gontcho A Gontcho}, S. and {Gonzalez}, D. and {Gonzalez-Morales}, A.~X. and {Gonzalez-Perez}, V. and {Gordon}, C. and {Graur}, O. and {Green}, D. and {Gruen}, D. and {Gsponer}, R. and {Guandalin}, C. and {Gutierrez}, G. and {Guy}, J. and {Hahn}, C. and {Han}, J.~J. and {Han}, J. and {He}, S. and {Herrera-Alcantar}, H.~K. and {Heydenreich}, S. and {Honscheid}, K. and {Hou}, J. and {Howlett}, C. and {Huterer}, D. and {Ir{\v{s}}i{\v{c}}}, V. and {Ishak}, M. and {Jacques}, A. and {Jiang}, L. and {Jimenez}, J. and {Jing}, Y.~P. and {Joachimi}, B. and {Joudaki}, S. and {Joyce}, R. and {Jullo}, E. and {Juneau}, S. and {Kara{\c{c}}ayl{\i}}, N.~G. and {Karim}, T. and {Kehoe}, R. and {Kent}, S. and {Khederlarian}, A. and {Kirkby}, D. and {Kisner}, T. and {Kitaura}, F.-S. and {Kizhuprakkat}, N. and {Kong}, H. and {Koposov}, S.~E. and {Kremin}, A. and {Krolewski}, A. and {Lahav}, O. and {Lai}, Y. and {Lamman}, C. and {Lan}, T.-W. and {Landriau}, M. and {Lang}, D. and {Lange}, J.~U. and {Lasker}, J. and {Le Goff}, J.~M. and {Le Guillou}, L. and {Leauthaud}, A. and {Levi}, M.~E. and {Li}, S. and {Li}, T.~S. and {Liu}, W. and {Lodha}, K. and {Lokken}, M. and {Luo}, Y. and {Magneville}, C. and {Manera}, M. and {Manser}, C.~J. and {Margala}, D. and {Martini}, P. and {Maus}, M. and {McCullough}, J. and {McDonald}, P. and {Medina}, G.~E. and {Medina-Varela}, L. and {Meisner}, A. and {Mena-Fern{\'a}ndez}, J. and {Menegas}, A. and {Meneses-Rizo}, J. and {Mezcua}, M. and {Miquel}, R. and {Montero-Camacho}, P. and {Moon}, J. and {Moustakas}, J. and {Mu{\~n}oz-Guti{\'e}rrez}, A. and {Mu noz-Santos}, D. and {Myers}, A.~D. and {Myles}, J. and {Nadathur}, S. and {Najita}, J. and {Napolitano}, L. and {Newman}, J.~A. and {Nikakhtar}, F. and {Nikutta}, R. and {Niz}, G. and {Noriega}, H.~E. and {Nugent}, P.},
        title = "{Data Release 1 of the Dark Energy Spectroscopic Instrument}",
      journal = {\aj},
         year = 2026,
        month = may,
       volume = {171},
       number = {5},
          eid = {285},
        pages = {285},
          doi = {10.3847/1538-3881/ae4c43},
archivePrefix = {arXiv},
       eprint = {2503.14745},
 primaryClass = {astro-ph.CO},
       adsurl = {https://ui.adsabs.harvard.edu/abs/2026AJ....171..285D}
}

@ARTICLE{Dieleman2015,
       author = {{Dieleman}, Sander and {Willett}, Kyle W. and {Dambre}, Joni},
        title = "{Rotation-invariant convolutional neural networks for galaxy morphology prediction}",
      journal = {\mnras},
         year = 2015,
        month = jun,
       volume = {450},
       number = {2},
        pages = {1441-1459},
          doi = {10.1093/mnras/stv632},
archivePrefix = {arXiv},
       eprint = {1503.07077},
 primaryClass = {astro-ph.IM},
       adsurl = {https://ui.adsabs.harvard.edu/abs/2015MNRAS.450.1441D}
}

@ARTICLE{Disney1976,
       author = {{Disney}, M.~J.},
        title = "{Visibility of galaxies}",
      journal = {\nat},
         year = 1976,
        month = oct,
       volume = {263},
       number = {5578},
        pages = {573-575},
          doi = {10.1038/263573a0},
       adsurl = {https://ui.adsabs.harvard.edu/abs/1976Natur.263..573D}
}

@ARTICLE{DominguezSanchez2018,
       author = {{Dom{\'\i}nguez S{\'a}nchez}, H. and {Huertas-Company}, M. and {Bernardi}, M. and {Tuccillo}, D. and {Fischer}, J.~L.},
        title = "{Improving galaxy morphologies for SDSS with Deep Learning}",
      journal = {\mnras},
         year = 2018,
        month = feb,
       volume = {476},
       number = {3},
        pages = {3661-3676},
          doi = {10.1093/mnras/sty338},
archivePrefix = {arXiv},
       eprint = {1711.05744},
 primaryClass = {astro-ph.GA},
       adsurl = {https://ui.adsabs.harvard.edu/abs/2018MNRAS.476.3661D}
}

@ARTICLE{Greco2018,
       author = {{Greco}, Johnny P. and {Greene}, Jenny E. and {Strauss}, Michael A. and {Macarthur}, Lauren A. and {Flowers}, Xzavier and {Goulding}, Andy D. and {Huang}, Song and {Kim}, Ji Hoon and {Komiyama}, Yutaka and {Leauthaud}, Alexie and {Leisman}, Lukas and {Lupton}, Robert H. and {Sif{\'o}n}, Crist{\'o}bal and {Wang}, Shiang-Yu},
        title = "{Illuminating Low Surface Brightness Galaxies with the Hyper Suprime-Cam Survey}",
      journal = {\apj},
         year = 2018,
        month = apr,
       volume = {857},
       number = {2},
          eid = {104},
        pages = {104},
          doi = {10.3847/1538-4357/aab842},
archivePrefix = {arXiv},
       eprint = {1709.04474},
 primaryClass = {astro-ph.GA},
       adsurl = {https://ui.adsabs.harvard.edu/abs/2018ApJ...857..104G}
}

@misc{Howard2017,
      title={MobileNets: Efficient Convolutional Neural Networks for Mobile Vision Applications}, 
      author={Andrew G. Howard and Menglong Zhu and Bo Chen and Dmitry Kalenichenko and Weijun Wang and Tobias Weyand and Marco Andreetto and Hartwig Adam},
      year={2017},
      eprint={1704.04861},
      archivePrefix={arXiv},
      primaryClass={cs.CV},
      url={https://arxiv.org/abs/1704.04861}, 
}

@misc{Huang2026unagi,
    author       = {Huang, Song},
    title        = {unagi: A Python client for the Hyper Suprime-Cam Subaru Strategic Program},
    year         = {2025},
    howpublished = {\url{https://github.com/dr-guangtou/unagi}},
    note         = {Accessed: 13 February 2026}
}

@article{HuertasCompany2015,
doi = {10.1088/0067-0049/221/1/8},
url = {https://doi.org/10.1088/0067-0049/221/1/8},
year = {2015},
month = {oct},
publisher = {The American Astronomical Society},
volume = {221},
number = {1},
pages = {8},
author = {Huertas-Company, M. and Gravet, R. and Cabrera-Vives, G. and Pérez-González, P. G. and Kartaltepe, J. S. and Barro, G. and Bernardi, M. and Mei, S. and Shankar, F. and Dimauro, P. and Bell, E. F. and Kocevski, D. and Koo, D. C. and Faber, S. M. and Mcintosh, D. H.},
title = {A CATALOG OF VISUAL-LIKE MORPHOLOGIES IN THE 5 CANDELS FIELDS USING DEEP LEARNING},
journal = {The Astrophysical Journal Supplement Series}
}

@article{Impey1997,
   author = "Impey, Chris and Bothun, Greg",
   title = "LOW SURFACE BRIGHTNESS GALAXIES", 
   journal= "Annual Review of Astronomy and Astrophysics",
   year = "1997",
   volume = "35",
   number = "Volume 35, 1997",
   pages = "267-307",
   doi = "https://doi.org/10.1146/annurev.astro.35.1.267",
   url = "https://www.annualreviews.org/content/journals/10.1146/annurev.astro.35.1.267",
   publisher = "Annual Reviews",
   issn = "1545-4282",
   type = "Journal Article",
  }

@ARTICLE{Impey1996,
       author = {{Impey}, C.~D. and {Sprayberry}, D. and {Irwin}, M.~J. and {Bothun}, G.~D.},
        title = "{Low Surface Brightness Galaxies in the Local Universe. I. The Catalog}",
      journal = {\apjs},
         year = 1996,
        month = aug,
       volume = {105},
        pages = {209},
          doi = {10.1086/192313},
       adsurl = {https://ui.adsabs.harvard.edu/abs/1996ApJS..105..209I}
}

@article{Jackson2021,
    author = {Jackson, R A and Martin, G and Kaviraj, S and Ramsøy, M and Devriendt, J E G and Sedgwick, T and Laigle, C and Choi, H and Beckmann, R S and Volonteri, M and Dubois, Y and Pichon, C and Yi, S K and Slyz, A and Kraljic, K and Kimm, T and Peirani, S and Baldry, I},
    title = {The origin of low-surface-brightness galaxies in the dwarf regime},
    journal = {Monthly Notices of the Royal Astronomical Society},
    volume = {502},
    number = {3},
    pages = {4262-4276},
    year = {2021},
    month = {01},
    issn = {0035-8711},
    doi = {10.1093/mnras/stab077},
    url = {https://doi.org/10.1093/mnras/stab077},
    eprint = {https://academic.oup.com/mnras/article-pdf/502/3/4262/38869818/stab077.pdf},
}

@ARTICLE{Kauffmann1993,
       author = {{Kauffmann}, G. and {White}, S.~D.~M. and {Guiderdoni}, B.},
        title = "{The formation and evolution of galaxies within merging dark matter haloes.}",
      journal = {\mnras},
         year = 1993,
        month = sep,
       volume = {264},
        pages = {201-218},
          doi = {10.1093/mnras/264.1.201},
       adsurl = {https://ui.adsabs.harvard.edu/abs/1993MNRAS.264..201K}
}

@article{Liang2024AutomaticLearning,
doi = {10.3847/1538-3881/ad4f8a},
url = {https://doi.org/10.3847/1538-3881/ad4f8a},
year = {2024},
month = {jul},
publisher = {The American Astronomical Society},
volume = {168},
number = {2},
pages = {74},
author = {Liang, Zengxu and Yi, Zhenping and Du, Wei and Liu, Meng and Liu, Yuan and Wang, Junjie and Kong, Xiaoming and Bu, Yude and Su, Hao and Wu, Hong},
title = {Automatic Search for Low-surface-brightness Galaxies from Sloan Digital Sky Survey Images Using Deep Learning},
journal = {The Astronomical Journal}
}

@misc{Loshchilov2019DecoupledRegularization,
      title={Decoupled Weight Decay Regularization}, 
      author={Ilya Loshchilov and Frank Hutter},
      year={2019},
      eprint={1711.05101},
      archivePrefix={arXiv},
      primaryClass={cs.LG},
      url={https://arxiv.org/abs/1711.05101}, 
}

@ARTICLE{Martin2019,
       author = {{Martin}, G. and {Kaviraj}, S. and {Laigle}, C. and {Devriendt}, J.~E.~G. and {Jackson}, R.~A. and {Peirani}, S. and {Dubois}, Y. and {Pichon}, C. and {Slyz}, A.},
        title = "{The formation and evolution of low-surface-brightness galaxies}",
      journal = {\mnras},
         year = 2019,
        month = may,
       volume = {485},
       number = {1},
        pages = {796-818},
          doi = {10.1093/mnras/stz356},
archivePrefix = {arXiv},
       eprint = {1902.04580},
 primaryClass = {astro-ph.GA},
       adsurl = {https://ui.adsabs.harvard.edu/abs/2019MNRAS.485..796M}
}

@ARTICLE{Mihos2015,
       author = {{Mihos}, J. Christopher and {Durrell}, Patrick R. and {Ferrarese}, Laura and {Feldmeier}, John J. and {C{\^o}t{\'e}}, Patrick and {Peng}, Eric W. and {Harding}, Paul and {Liu}, Chengze and {Gwyn}, Stephen and {Cuillandre}, Jean-Charles},
        title = "{Galaxies at the Extremes: Ultra-diffuse Galaxies in the Virgo Cluster}",
      journal = {\apjl},
         year = 2015,
        month = aug,
       volume = {809},
       number = {2},
          eid = {L21},
        pages = {L21},
          doi = {10.1088/2041-8205/809/2/L21},
archivePrefix = {arXiv},
       eprint = {1507.02270},
 primaryClass = {astro-ph.GA},
       adsurl = {https://ui.adsabs.harvard.edu/abs/2015ApJ...809L..21M}
}

@ARTICLE{Moore1999,
       author = {{Moore}, B. and {Quinn}, T. and {Governato}, F. and {Stadel}, J. and {Lake}, G.},
        title = "{Cold collapse and the core catastrophe}",
      journal = {\mnras},
         year = 1999,
        month = dec,
       volume = {310},
       number = {4},
        pages = {1147-1152},
          doi = {10.1046/j.1365-8711.1999.03039.x},
archivePrefix = {arXiv},
       eprint = {astro-ph/9903164},
 primaryClass = {astro-ph},
       adsurl = {https://ui.adsabs.harvard.edu/abs/1999MNRAS.310.1147M}
}

@ARTICLE{Morishita2021SuperBoRG:Data,
       author = {{Morishita}, T.},
        title = "{SuperBoRG: Search for the Brightest of Reionizing Galaxies and Quasars in HST Parallel Imaging Data}",
      journal = {\apjs},
         year = 2021,
        month = mar,
       volume = {253},
       number = {1},
          eid = {4},
        pages = {4},
          doi = {10.3847/1538-4365/abce67},
archivePrefix = {arXiv},
       eprint = {2010.15637},
 primaryClass = {astro-ph.GA},
       adsurl = {https://ui.adsabs.harvard.edu/abs/2021ApJS..253....4M}
}

@article{Ntampaka2019,
	author = {Ntampaka, Michelle and Avestruz, Camille and Boada, Steven and Caldeira, Joao and Cisewski-Kehe, Jessi and Stefano, Rosanne Di and Dvorkin, Cora and Evrard, August E. and Farahi, Arya and Finkbeiner, Doug and Genel, Shy and Goodman, Alyssa and Goulding, Andy and Ho, Shirley and Kosowsky, Arthur and Plante, Paul La and Lanusse, Francois and Lochner, Michelle and Mandelbaum, Rachel and Nagai, Daisuke and Newman, Jeffrey A. and Nord, Brian and Peek, J. E. G. and Peel, Austin and Poczos, Barnabas and Rau, Markus Michael and Siemiginowska, Aneta and Sutherland, Dougal J. and Trac, Hy and Wandelt, Benjamin},
	journal = {Bulletin of the AAS},
	number = {3},
	year = {2019},
	month = {may 31},
	note = {https://baas.aas.org/pub/2020n3i014},
	publisher = {},
	title = {The {Role} of {Machine} {Learning} in the {Next} {Decade} of {Cosmology}},
	volume = {51},
}

@software{pygalfitm_software,
  author       = {Oliveira Schwarz, G. B. and
                  Cortesi, A. and
                  Okiyama, L.},
  title        = {pygalfitm},
  month        = feb,
  year         = 2023,
  publisher    = {Zenodo},
  doi          = {10.5281/zenodo.15648590},
  url          = {https://doi.org/10.5281/zenodo.15648590}
}

@ARTICLE{galfitm_software,
       author = {{Vika}, Marina and {Bamford}, Steven P. and {H{\"a}u{\ss}ler}, Boris and {Rojas}, Alex L. and {Borch}, Andrea and {Nichol}, Robert C.},
        title = "{MegaMorph - multiwavelength measurement of galaxy structure. S{\'e}rsic profile fits to galaxies near and far}",
      journal = {\mnras},
         year = 2013,
        month = oct,
       volume = {435},
       number = {1},
        pages = {623-649},
          doi = {10.1093/mnras/stt1320},
archivePrefix = {arXiv},
       eprint = {1307.4996},
 primaryClass = {astro-ph.CO},
       adsurl = {https://ui.adsabs.harvard.edu/abs/2013MNRAS.435..623V}
}

@ARTICLE{Peng2002,
       author = {{Peng}, Chien Y. and {Ho}, Luis C. and {Impey}, Chris D. and {Rix}, Hans-Walter},
        title = "{Detailed Structural Decomposition of Galaxy Images}",
      journal = {\aj},
         year = 2002,
        month = jul,
       volume = {124},
       number = {1},
        pages = {266-293},
          doi = {10.1086/340952},
archivePrefix = {arXiv},
       eprint = {astro-ph/0204182},
 primaryClass = {astro-ph},
       adsurl = {https://ui.adsabs.harvard.edu/abs/2002AJ....124..266P}
}

@article{Peng2010,
doi = {10.1088/0004-6256/139/6/2097},
url = {https://doi.org/10.1088/0004-6256/139/6/2097},
year = {2010},
month = {apr},
publisher = {The American Astronomical Society},
volume = {139},
number = {6},
pages = {2097},
author = {Peng, Chien Y. and Ho, Luis C. and Impey, Chris D. and Rix, Hans-Walter},
title = {DETAILED DECOMPOSITION OF GALAXY IMAGES. II. BEYOND AXISYMMETRIC MODELS},
journal = {The Astronomical Journal}
}

@article{Prole2019,
    author = {Prole, D J and Hilker, M and van der Burg, R F J and Cantiello, M and Venhola, A and Iodice, E and van de Ven, G and Wittmann, C and Peletier, R F and Mieske, S and Capaccioli, M and Napolitano, N R and Paolillo, M and Spavone, M and Valentijn, E},
    title = {Halo mass estimates from the globular cluster populations of 175 low surface brightness galaxies in the Fornax cluster},
    journal = {Monthly Notices of the Royal Astronomical Society},
    volume = {484},
    number = {4},
    pages = {4865-4880},
    year = {2019},
    month = {04},
    issn = {0035-8711},
    doi = {10.1093/mnras/stz326},
    url = {https://doi.org/10.1093/mnras/stz326},
    eprint = {https://academic.oup.com/mnras/article-pdf/484/4/4865/27781986/stz326.pdf},
}

@ARTICLE{Perez-Montano2022,
       author = {{P{\'e}rez-Monta{\~n}o}, Luis Enrique and {Rodriguez-Gomez}, Vicente and {Cervantes Sodi}, Bernardo and {Zhu}, Qirong and {Pillepich}, Annalisa and {Vogelsberger}, Mark and {Hernquist}, Lars},
        title = "{The formation of low surface brightness galaxies in the IllustrisTNG simulation}",
      journal = {\mnras},
         year = 2022,
        month = aug,
       volume = {514},
       number = {4},
        pages = {5840-5852},
          doi = {10.1093/mnras/stac1716},
archivePrefix = {arXiv},
       eprint = {2206.08942},
 primaryClass = {astro-ph.GA},
       adsurl = {https://ui.adsabs.harvard.edu/abs/2022MNRAS.514.5840P}
}

@ARTICLE{Rosenbaum2009,
       author = {{Rosenbaum}, S.~D. and {Krusch}, E. and {Bomans}, D.~J. and {Dettmar}, R.-J.},
        title = "{The large-scale environment of low surface brightness galaxies}",
      journal = {\aap},
         year = 2009,
        month = sep,
       volume = {504},
       number = {3},
        pages = {807-820},
          doi = {10.1051/0004-6361/20077462},
archivePrefix = {arXiv},
       eprint = {0908.4023},
 primaryClass = {astro-ph.CO},
       adsurl = {https://ui.adsabs.harvard.edu/abs/2009A&A...504..807R}
}

@article{Stoppacher2025,
	author = {{Stoppacher, D.} and {Tissera, P.} and {Rosas-Guevara, Y.} and {Galaz, G.} and {Oñorbe, J.}},
	title = {Hidden figures in the sky - Evolution of low-surface-brightness galaxies from a hydrodynamical perspective},
	DOI= "10.1051/0004-6361/202555232",
	url= "https://doi.org/10.1051/0004-6361/202555232",
	journal = {A\&A},
	year = 2025,
	volume = 701,
	pages = "A272",
}

@ARTICLE{Su2024LSBGnet:Galaxies,
       author = {{Su}, Hao and {Yi}, Zhenping and {Liang}, Zengxu and {Du}, Wei and {Liu}, Meng and {Kong}, Xiaoming and {Bu}, Yude and {Wu}, Hong},
        title = "{LSBGnet: an improved detection model for low-surface brightness galaxies}",
      journal = {\mnras},
         year = 2024,
        month = feb,
       volume = {528},
       number = {1},
        pages = {873-882},
          doi = {10.1093/mnras/stae001},
       adsurl = {https://ui.adsabs.harvard.edu/abs/2024MNRAS.528..873S}
}

@article{ Su:2026:A&A,
	author = {{Su, Hao} and {Li, Rui} and {Napolitano, Nicola R.} and {Yi, Zhenping} and {Tortora, Crescenzo} and {Su, Yiping} and {Kuijken, Konrad} and {Chen, Liqing} and {Li, Ran} and {Ragusa, Rossella} and {Li, Sihan} and {Dong, Yue} and {Radovich, Mario} and {Wright, Angus H.} and {Covone, Giovanni} and {Zhong, Fucheng}},
	title = {Ultra-diffuse galaxies in the Kilo-Degree Survey with deep learning},
	DOI= "10.1051/0004-6361/202556535",
	url= "https://doi.org/10.1051/0004-6361/202556535",
	journal = {A\&A},
	year = 2026,
	volume = 707,
	pages = "A354",
}

@article{Tanoglidis2021,
doi = {10.3847/1538-4365/abca89},
url = {https://doi.org/10.3847/1538-4365/abca89},
year = {2021},
month = {jan},
publisher = {The American Astronomical Society},
volume = {252},
number = {2},
pages = {18},
author = {Tanoglidis, D. and Drlica-Wagner, A. and Wei, K. and Li, T. S. and Sánchez, J. and Zhang, Y. and Peter, A. H. G. and Feldmeier-Krause, A. and Prat, J. and Casey, K. and Palmese, A. and Sánchez, C. and DeRose, J. and Conselice, C. and Gagnon, L. and Abbott, T. M. C. and Aguena, M. and Allam, S. and Avila, S. and Bechtol, K. and Bertin, E. and Bhargava, S. and Brooks, D. and Burke, D. L. and Rosell, A. Carnero and Kind, M. Carrasco and Carretero, J. and Chang, C. and Costanzi, M. and Costa, L. N. da and Vicente, J. De and Desai, S. and Diehl, H. T. and Doel, P. and Eifler, T. F. and Everett, S. and Evrard, A. E. and Flaugher, B. and Frieman, J. and García-Bellido, J. and Gerdes, D. W. and Gruendl, R. A. and Gschwend, J. and Gutierrez, G. and Hartley, W. G. and Hollowood, D. L. and Huterer, D. and James, D. J. and Krause, E. and Kuehn, K. and Kuropatkin, N. and Maia, M. A. G. and March, M. and Marshall, J. L. and Menanteau, F. and Miquel, R. and Ogando, R. L. C. and Paz-Chinchón, F. and Romer, A. K. and Roodman, A. and Sanchez, E. and Scarpine, V. and Serrano, S. and Sevilla-Noarbe, I. and Smith, M. and Suchyta, E. and Tarle, G. and Thomas, D. and Tucker, D. L. and Walker, A. R. and (DES Collaboration)},
title = {Shadows in the Dark: Low-surface-brightness Galaxies Discovered in the Dark Energy Survey},
journal = {The Astrophysical Journal Supplement Series}
}

@article{ Thuruthipilly2025DESLearning,
	author = {{Thuruthipilly, H.} and {Junais} and {Koda, J.} and {Pollo, A.} and {Yagi, M.} and {Yamanoi, H.} and {Komiyama, Y.} and {Romano, M.} and {Małek, K.} and {Donevski, D.}},
	title = {DES to HSC: Detecting low-surface-brightness galaxies in the Abell 194 cluster using transfer learning},
	DOI= "10.1051/0004-6361/202452934",
	url= "https://doi.org/10.1051/0004-6361/202452934",
	journal = {A\&A},
	year = 2025,
	volume = 695,
	pages = "A106",
}

@ARTICLE{KIDS,
       author = {{Wright}, Angus H. and {Kuijken}, Konrad and {Hildebrandt}, Hendrik and {Radovich}, Mario and {Bilicki}, Maciej and {Dvornik}, Andrej and {Getman}, Fedor and {Heymans}, Catherine and {Hoekstra}, Henk and {Li}, Shun-Sheng and {Miller}, Lance and {Napolitano}, Nicola R. and {Xia}, Qianli and {Asgari}, Marika and {Brescia}, Massimo and {Buddelmeijer}, Hugo and {Burger}, Pierre and {Castignani}, Gianluca and {Cavuoti}, Stefano and {de Jong}, Jelte and {Edge}, Alastair and {Giblin}, Benjamin and {Giocoli}, Carlo and {Harnois-D{\'e}raps}, Joachim and {Jalan}, Priyanka and {Joachimi}, Benjamin and {John William}, Anjitha and {Joudaki}, Shahab and {Kannawadi}, Arun and {Kaur}, Gursharanjit and {La Barbera}, Francesco and {Linke}, Laila and {Mahony}, Constance and {Maturi}, Matteo and {Moscardini}, Lauro and {Nakoneczny}, Szymon J. and {Paolillo}, Maurizio and {Porth}, Lucas and {Puddu}, Emanuella and {Reischke}, Robert and {Schneider}, Peter and {Sereno}, Mauro and {Shan}, HuanYuan and {Sif{\'o}n}, Crist{\'o}bal and {St{\"o}lzner}, Benjamin and {Tr{\"o}ster}, Tilman and {Valentijn}, Edwin and {van den Busch}, Jan Luca and {Verdoes Kleijn}, Gijs and {Wittje}, Anna and {Yan}, Ziang and {Yao}, Ji and {Yoon}, Mijin and {Zhang}, Yun-Hao},
        title = "{The fifth data release of the Kilo Degree Survey: Multi-epoch optical/NIR imaging covering wide and legacy-calibration fields}",
      journal = {\aap},
         year = 2024,
        month = jun,
       volume = {686},
          eid = {A170},
        pages = {A170},
          doi = {10.1051/0004-6361/202346730},
archivePrefix = {arXiv},
       eprint = {2503.19439},
 primaryClass = {astro-ph.GA},
       adsurl = {https://ui.adsabs.harvard.edu/abs/2024A&A...686A.170W}
}

@ARTICLE{Zhong2008,
       author = {{Zhong}, G.~H. and {Liang}, Y.~C. and {Liu}, F.~S. and {Hammer}, F. and {Hu}, J.~Y. and {Chen}, X.~Y. and {Deng}, L.~C. and {Zhang}, B.},
        title = "{A large sample of low surface brightness disc galaxies from the SDSS - I. The sample and the stellar populations}",
      journal = {\mnras},
         year = 2008,
        month = dec,
       volume = {391},
       number = {2},
        pages = {986-999},
          doi = {10.1111/j.1365-2966.2008.13972.x},
archivePrefix = {arXiv},
       eprint = {0809.3099},
 primaryClass = {astro-ph},
       adsurl = {https://ui.adsabs.harvard.edu/abs/2008MNRAS.391..986Z}
}

\end{document}